\documentclass[pdflatex,sn-mathphys-num,iicol]{sn-jnl}

\usepackage{graphicx}
\usepackage{multirow}
\usepackage{amsmath,amssymb,amsfonts}
\usepackage{amsthm}
\usepackage{mathrsfs}
\usepackage[title]{appendix}
\usepackage{xcolor}
\usepackage{textcomp}
\usepackage{manyfoot}
\usepackage{booktabs}
\usepackage{capt-of}
\usepackage[left]{lineno}

\begin{document}

\title[Noise-aware derivative feedback under quantum-limited measurement]
{Noise-aware derivative feedback of a harmonic oscillator under quantum-limited position measurement}

\author*[1,2]{\fnm{Vedran} \sur{Vujnović}}\email{vedran.vujnovic@uniri.hr}

\affil[1]{Medical Physics and Radiation Protection Department, University Hospital Rijeka, 51000 Rijeka, Croatia}
\affil[2]{Laboratory for Quantum and Nonlinear Optics, Faculty of Physics and Centre for Micro- and Nanosciences and Technologies, University of Rijeka, Radmile Matejčić 2, 51000 Rijeka, Croatia}

\abstract{
We study measurement-based feedback control of a damped harmonic oscillator, motivated by optomechanical and interferometric position measurements, in which the feedback force is synthesized directly from a displacement measurement.
For a band-limited derivative (``dirty-derivative'') controller, the feedback loop simultaneously shapes the mechanical susceptibility and reinjects measurement noise back onto the oscillator, creating a nontrivial tradeoff between stabilization and noise amplification.
We first present a compact classical formulation that isolates this tradeoff in the closed-loop displacement spectrum and identifies the controller parameters that determine resonance suppression versus noise-driven actuation.
We then extend the same topology to a quantum-limited position detector, enforcing the imprecision--backaction constraint and introducing a finite measurement bandwidth that both regularizes high-frequency noise reinjection and adds phase lag in the feedback path.
In the high-$Q$ regime we obtain a compact analytic design rule for the optimal feedback gain at fixed controller and measurement bandwidths and compare it directly with full-spectrum calculations.
The comparison shows that a high mechanical quality factor alone does not guarantee quantitative agreement: the accuracy of the near-resonant approximation also depends strongly on the controller and measurement bandwidths.
Because finite measurement bandwidth changes both the magnitude and phase of the feedback loop, the analysis separates the near-resonant damping contribution from imprecision-noise reinjection and shows that positive near-resonant feedback damping requires the product of the controller cutoff and measurement bandwidth to exceed the square of the mechanical resonance frequency.
Finally, we present a dimensionless map of the gain-optimized high-$Q$ occupation and compare its predictions with full-spectrum optimization.
All analytical gains evaluated in this map satisfy the full closed-loop stability criterion for the stated parameters.
}


\maketitle
\section{Introduction}

Measurement-based feedback is widely used to stabilize mechanical degrees of freedom, suppress resonances, and shape effective dynamics in precision experiments.
Examples include cavity optomechanics, nanomechanical and electromechanical resonators, AFM cantilevers, interferometric displacement sensors, and inertial reference systems.
Across these systems, a common ``direct-output'' topology appears: a mechanical displacement is continuously measured and an actuation force synthesized from the measurement record is fed back to the same mode.
In this topology the loop modifies the mechanical susceptibility while inevitably reinjecting measurement noise through the actuation path.
Reviews and tutorials covering this cross-platform viewpoint include Refs.~\cite{bechhoefer2021control,aastrom2021feedback,AspelmeyerRMP2014,ClerkRMP2010}.

In this paper, the oscillator is treated at the level of an effective mechanical mode under continuous position readout -- optomechanical systems provide one important realization, but the transfer-function description also applies to other quantum-limited displacement sensors.

Here we focus on band-limited derivative feedback, often called cold damping. 
Derivative feedback increases the effective damping of the targeted mode and suppresses resonant motion, while a finite derivative roll-off limits high-frequency noise amplification. 
Cold-damping feedback has been studied in early interferometric and optomechanical experiments \cite{CohadonPRL1999,PinardPRA2000,CourtyHeidmannPinard2001} and subsequently in cantilever and force-sensing systems \cite{PoggioPRL2007}.
Experiments have progressively extended measurement-based mechanical control toward the quantum regime \cite{WilsonNature2015,GuoPRL2019,RossiNature2018,MagriniNature2021,GuoNatCommun2023}, while coherent-feedback schemes provide a complementary route that avoids an explicit measurement record \cite{ErnzerPRX2023}.

The basic ingredients of the problem are well established. 
Measurement-based quantum feedback has a long history in continuous-measurement theory \cite{WisemanMilburn1993,Wiseman1994,WisemanMilburn2010,Jacobs2014}, while quantum-limited linear position detection and its imprecision--backaction constraint are reviewed in Ref.~\cite{ClerkRMP2010}.
Cold damping of mechanical motion has likewise been demonstrated and analyzed in interferometric and optomechanical settings \cite{CohadonPRL1999,PinardPRA2000,CourtyHeidmannPinard2001,GenesPRA2008,AspelmeyerRMP2014}.
The focus here is narrower and concerns the joint effect of a finite derivative-controller bandwidth and a finite measurement bandwidth in direct measurement-based feedback, where the same measured record both supplies the damping signal and carries the imprecision noise that is fed back onto the oscillator.

The contribution of the present analysis is to separate the resulting finite-bandwidth response into its near-resonant damping and imprecision-reinjection components.
We show that useful feedback damping and imprecision-noise reinjection are governed by distinct factors, $\alpha_d$ and $\alpha_n$, so that finite bandwidth does not affect damping and noise reinjection in the same way.
This separation yields the condition for positive near-resonant feedback damping $\omega_f\omega_{\rm meas}>\omega_0^2$, a compact high-$Q$ expression for the optimal feedback gain, and a dimensionless design map in $(\omega_f/\omega_0,\omega_{\rm meas}/\omega_0)$.
The near-resonant design rule is then tested directly against the full-spectrum calculation, showing that its quantitative accuracy depends not only on the mechanical quality factor but also on the controller and measurement bandwidths.

We derive the stability criterion of the full closed-loop dynamics and verify that all high-$Q$ gains evaluated in the design map satisfy it for the stated parameters.
A separate full-spectrum optimization quantifies the error of the predicted minimum occupation across the same bandwidth plane, identifying regions of agreement within $10\%$ and $25\%$.
The largest discrepancies occur for broad, approximately equal bandwidths, where stable analytical gains can nevertheless produce substantially suboptimal full-spectrum occupations.

The paper is organized as follows. 
Section~\ref{sec:classical_minimal} presents the classical feedback topology and noise-reinjection mechanism.
Section~\ref{sec:quantum_feedback} extends the same topology to a quantum-limited detector, derives the occupation-based figure of merit, and presents the finite-bandwidth and high-$Q$ design results. 
The appendices give the feedback conventions, discuss correlated detector noise, derive the high-$Q$ occupation estimate, summarize the notation, establish the full closed-loop stability criterion, and derive the full-frequency moments used for gain optimization.

\section{Classical loop and noise reinjection}
\label{sec:classical_minimal}

We consider a harmonic oscillator driven by external forcing and an added feedback force,
\begin{equation}
\ddot{x}(t) + 2\gamma_u \dot{x}(t) + \omega_0^2 x(t) = \frac{1}{m}\!\left[F_{\rm ext}(t)+F_{\rm fb}(t)\right],
\label{eq:classical_osc}
\end{equation}
Here $\gamma_u$ denotes the intrinsic mechanical amplitude-decay rate; with the convention of Eq.~\eqref{eq:classical_osc}, $2\gamma_u$ is the energy-damping rate and
$Q=\omega_0/(2\gamma_u)$.
The measured displacement record is modeled as
\begin{equation}
y(\omega)=x(\omega)+x_{\rm imp}(\omega),
\label{eq:classical_meas_record}
\end{equation}
where $x_{\rm imp}$ denotes additive measurement imprecision noise
\cite{ClerkRMP2010,JacobsSteck2006,WisemanMilburn2010}.

\paragraph*{Band-limited derivative force from the readout.} 
We synthesize the feedback force directly from the measured record (see Appendix~\ref{app:topology}). 
In the ideal derivative limit, the corresponding time-domain feedback law is
$
F_{\rm fb}(t)=-2m\gamma_{\rm fb}\,\dot{y}(t),
$
so the band-limited derivative controller (Appendix~\ref{app:filtered_derivative}) is therefore
\begin{equation}
F_{\rm fb}(\omega)=-G(\omega)\,y(\omega),
\quad
G(\omega)=\frac{2m\gamma_{\rm fb}\,i\omega}{1+i\omega/\omega_f}.
\label{eq:classical_dirty_derivative}
\end{equation}
Here $\gamma_{\rm fb}$ denotes the feedback-damping gain, distinct from the intrinsic rate $\gamma_u$, while the roll-off frequency $\omega_f$ bounds the high-frequency gain and limits amplification of measurement noise under the commonly used white-imprecision idealization \cite{bechhoefer2021control,aastrom2021feedback}. 
Throughout the paper, we use the notation $G(\omega)\equiv G(i\omega)$ for the frequency response obtained by evaluating the transfer function at $s=i\omega$.

\paragraph*{Closed-loop susceptibility and displacement spectrum.}
The open-loop mechanical susceptibility is
\begin{equation}
\chi_m(\omega)=\frac{1}{m\left(\omega_0^2-\omega^2+i\,2\gamma_u\omega\right)}.
\label{eq:classical_chi_m}
\end{equation}
Combining Eqs.~\eqref{eq:classical_osc}--\eqref{eq:classical_dirty_derivative} yields
\begin{equation}
x(\omega)=\chi_{\rm eff}(\omega)\Big[F_{\rm ext}(\omega)-G(\omega)\,x_{\rm imp}(\omega)\Big],
\end{equation}
\begin{equation}
\chi_{\rm eff}(\omega)=\frac{\chi_m(\omega)}{1+\chi_m(\omega)G(\omega)}.
\label{eq:classical_chi_eff}
\end{equation}
Thus, for external force noise with spectrum $S_{FF}^{\rm ext}(\omega)$ and imprecision spectrum $S_{xx}^{\rm imp}(\omega)$, the displacement spectrum is
\begin{equation}
S_{xx}(\omega)=|\chi_{\rm eff}(\omega)|^2\left(S_{FF}^{\rm ext}(\omega)+|G(\omega)|^2 S_{xx}^{\rm imp}(\omega)\right).
\label{eq:classical_Sxx}
\end{equation}
The measurement-noise transfer is also derived in Appendix~\ref{app:meas_noise_injection}.
Eq.~\eqref{eq:classical_Sxx} highlights the key tradeoff used throughout the paper: increasing $\gamma_{\rm fb}$ broadens the resonance (increased effective damping through $\chi_{\rm eff}$) but simultaneously reinjects measurement imprecision as an effective force noise proportional to $|G|^2 S_{xx}^{\rm imp}$.
In classical modeling, $S_{FF}^{\rm ext}$ and $S_{xx}^{\rm imp}$ may be treated as independent. 
However in the quantum-limited extension below, imprecision and backaction are constrained and this tradeoff produces a genuine optimum \cite{ClerkRMP2010}.

\paragraph*{Motivation for the quantum extension.}
The classical formulation above shows that derivative feedback competes against noise reinjection, and that finite bandwidth is essential to keep high-frequency contributions controlled. 
In the quantum-limited regime, this competition becomes unavoidable because measurement imprecision and backaction cannot be reduced independently \cite{ClerkRMP2010}. 

\section{Quantum-limited feedback}
\label{sec:quantum_feedback}

This section reformulates the same feedback topology for a quantum-limited position detector and addresses a focused controller-design question: for band-limited derivative feedback with finite measurement bandwidth, what gain and cutoff minimize the steady-state occupation, and in which regimes does increased derivative action worsen performance rather than improve stabilization?
We provide a quantum extension of the classical analysis by treating the readout as a \emph{quantum-limited position detector} and retaining the same measurement-based feedback topology used throughout this paper.
Continuous measurement and measurement-based feedback are standard in quantum control theory \cite{JacobsSteck2006,Wiseman1994,WisemanMilburn2010}, and the quantum-noise constraints for linear position detection can be expressed compactly in the linear-response framework \cite{ClerkRMP2010}.

\begin{figure*}[t]
\centering
\includegraphics[width=0.85\textwidth]{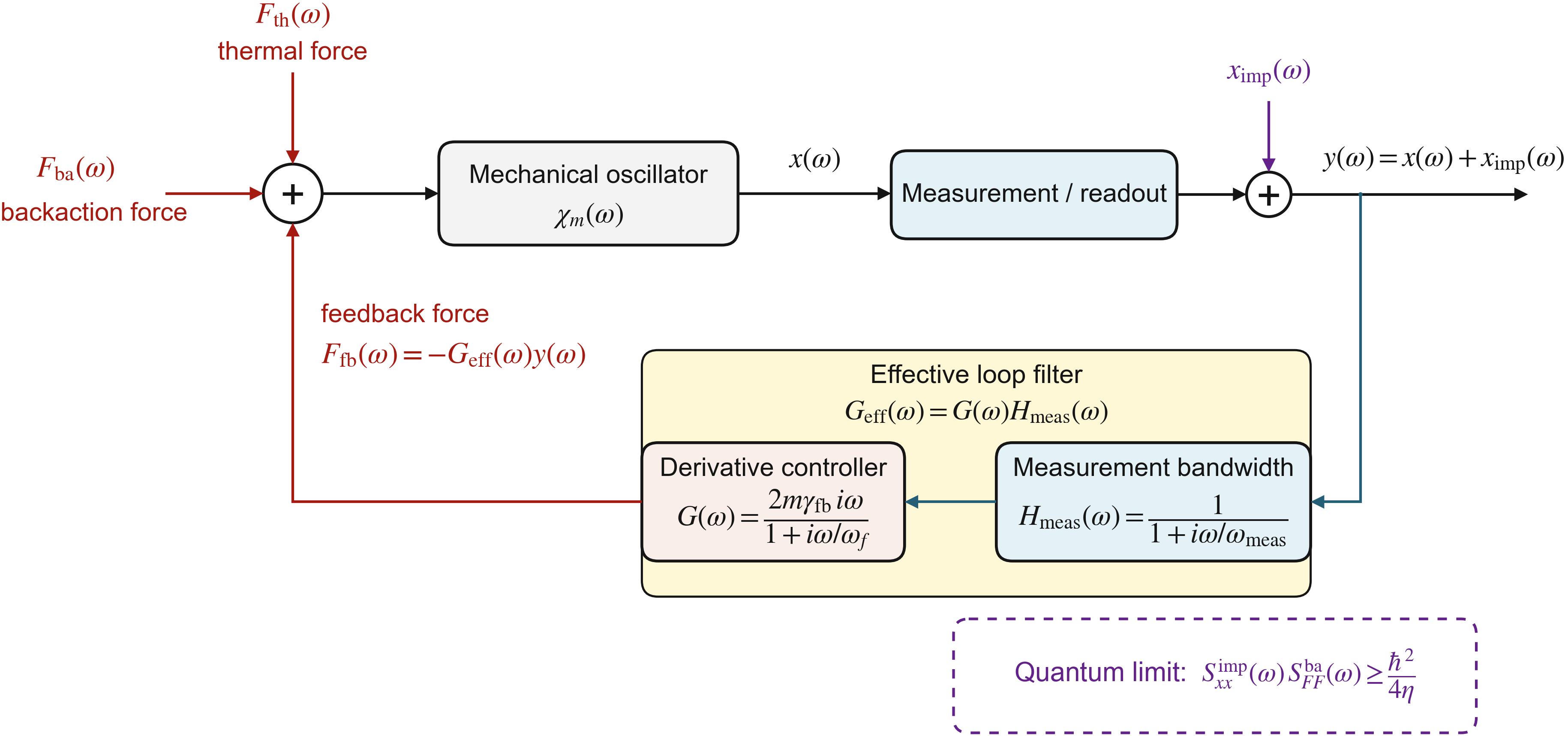}
\vspace{1mm}
  \caption{
  Schematic of the measurement-based derivative-feedback loop used in the quantum-limited analysis.
  Thermal force $F_{\rm th}$, measurement backaction $F_{\rm ba}$, and feedback force $F_{\rm fb}$ drive the mechanical oscillator with susceptibility $\chi_m(\omega)$.
  The measured record is $y(\omega)=x(\omega)+x_{\rm imp}(\omega)$, where $x_{\rm imp}$ is the imprecision noise.
  The feedback force is synthesized from the measured record through the effective loop filter $G_{\rm eff}(\omega)=G(\omega)H_{\rm meas}(\omega)$, giving  $F_{\rm fb}(\omega)=-G_{\rm eff}(\omega)y(\omega)$.
  The imprecision and backaction noises obey the quantum-limit constraint $S_{xx}^{\rm imp}(\omega)S_{FF}^{\rm ba}(\omega)\ge \hbar^2/(4\eta)$.
  }
  \label{fig:feedback_loop_schematic}
\end{figure*}

We keep the oscillator dynamics in classical linear time-invariant form (see Appendix~\ref{app:supp_classical_overview}) and incorporate quantum mechanics through the quantum-limited detector noise, namely imprecision and backaction, and through the occupation-based performance metric. 
This is equivalent to the standard linear-Gaussian continuous-measurement treatment.
The corresponding measurement-based feedback topology is shown schematically in Fig.~\ref{fig:feedback_loop_schematic}.

\subsection{Model and measurement}
We consider the harmonic oscillator model
\begin{equation}
\ddot{x}(t) + 2\gamma_u \dot{x}(t) + \omega_0^2 x(t) = \frac{1}{m}\!\left[F_{\rm th}(t) + F_{\rm ba}(t) + F_{\rm fb}(t)\right].
\label{eq:q_osc}
\end{equation}
A continuous position measurement yields a classical measurement record $y(\omega)$ as defined in Eq.\,\eqref{eq:classical_meas_record}. 
The measurement necessarily produces a backaction force noise $F_{\rm ba}$ acting on the oscillator.
For a quantum-limited detector with efficiency $\eta\in(0,1]$, the (symmetrized) noise spectra obey the constraint \cite{ClerkRMP2010}
\begin{equation}
S_{xx}^{\rm imp}(\omega)\,S_{FF}^{\rm ba}(\omega)\ge \frac{\hbar^2}{4\eta}.
\label{eq:quantum_limit}
\end{equation}
For all quantitative results below, we assume saturation of Eq.~\eqref{eq:quantum_limit}, $S_{xx}^{\rm imp}(\omega)S_{FF}^{\rm ba}(\omega)=\hbar^2/(4\eta)$, for every $\eta\in(0,1]$, and neglect imprecision--backaction correlations and additional classical noise sources.
Thus, within the model, detector inefficiency is represented solely through $\eta$.
Throughout, $S_{xx}^{\rm imp}$, $S_{FF}^{\rm ba}$, and $S_{FF}^{\rm th}$ denote two-sided symmetrized noise spectra.
The same convention is used for the resulting displacement spectrum $S_{xx}$.
Throughout the theoretical development, $\omega$, $\omega_0$, $\omega_f$, and $\omega_{\rm meas}$ denote angular frequencies.
When quantities denoted by $\omega$ or $\kappa$ are expressed in hertz, division by $2\pi$ is shown explicitly.
Accordingly, all quantitative results and figures presented in this section use the uncorrelated detector model.
Correlated imprecision--backaction noise is discussed separately in Appendix~\ref{app:quantum_correlations} as a natural extension of the present analysis.
Thermal forcing is modeled as Markovian near $\omega_0$:
\begin{equation}
S_{FF}^{\rm th}\simeq 4m\gamma_u\,\hbar\omega_0\!\left(n_{\rm th}+\frac12\right),
\label{eq:thermal_force}
\end{equation}
with
\begin{equation}
n_{\rm th}(T)=\frac{1}{e^{\hbar\omega_0/k_B T}-1}.
\label{eq:nth_def}
\end{equation}
Here $n_{\rm th}(T)$ is the mean thermal occupation at bath temperature $T$.
The classical limit is
$S_{FF}^{\rm th}\approx 4m\gamma_u k_B T$
when $k_BT\gg\hbar\omega_0$.
Thus, the temperature values used in the examples may be mapped to discrete $n_{\rm th}$ values via Eq.~\eqref{eq:nth_def}.
For the low-frequency choice $\omega_0=2~\mathrm{rad/s}$, laboratory temperatures correspond to very large $n_{\rm th}$, whereas the small-$n_{\rm th}$ regime illustrates the quantum-limited behavior.

\paragraph*{Band-limited derivative feedback.} 

We use a \emph{band-limited (filtered) derivative} controller, sometimes referred to as a ``dirty derivative'' in the control literature, to ensure bounded high-frequency gain (see Eq.~\eqref{eq:app_filtered_derivative} in Appendix~\ref{app:filtered_derivative}).
When acting on the measurement record,
\begin{equation}
F_{\rm fb}(\omega)=-G(\omega)\,y(\omega),
~~\,
G(\omega)=\frac{2m\gamma_{\rm fb}\,i\omega}{1+i\omega/\omega_f},
\label{eq:band-limited_derivative}
\end{equation}
where $\gamma_{\rm fb}$ sets the feedback-damping scale and $\omega_f$ is a finite roll-off frequency. 
Such filtering is essential in practice because ideal differentiation amplifies high-frequency measurement noise and can render steady-state variances ill-defined under a white-imprecision model. 
Physically, finite detector and actuator bandwidths impose an effective high-frequency cutoff \cite{GenesPRA2008,HabibiJOpt2016}.

\paragraph*{Closed-loop susceptibility and noise injection.}

Combining Eq.~\eqref{eq:q_osc} with the measurement record in Eq.~\eqref{eq:classical_meas_record} and the feedback law in Eq.~\eqref{eq:band-limited_derivative}, and using the open-loop mechanical susceptibility from Eq.~\eqref{eq:classical_chi_m}, yields the closed-loop response
\begin{equation}
x(\omega)
=
\chi_{\rm eff}(\omega)
\Big[
F_{\rm th}(\omega)+F_{\rm ba}(\omega)-G(\omega)x_{\rm imp}(\omega)
\Big],
\label{eq:quantum_closed_loop_response}
\end{equation}
where $\chi_{\rm eff}(\omega)$ is given by Eq.~\eqref{eq:classical_chi_eff}.
The displacement spectrum therefore becomes
\begin{equation}
S_{xx}(\omega)=|\chi_{\rm eff}(\omega)|^2\left(S_{FF}^{\rm th}+S_{FF}^{\rm ba}+|G(\omega)|^2 S_{xx}^{\rm imp}\right),
\label{eq:Sxx_closed}
\end{equation}
which makes explicit the tradeoff: increasing $\gamma_{\rm fb}$ broadens the resonance (larger effective damping) but also reinjects measurement imprecision as a force noise term proportional to $|G|^2S_{xx}^{\rm imp}$.
In contrast to classical modeling where $S_{xx}^{\rm imp}$ and $S_{FF}^{\rm ba}$ can be tuned independently, Eq.~\eqref{eq:quantum_limit} couples them and produces a genuine optimum.

\paragraph*{Figure of merit and non-monotonic feedback regime.}

A convenient steady-state figure of merit is the effective occupation number $n_{\rm eff}$ obtained from
\begin{align}
&\langle x^2\rangle
=
\frac{1}{2\pi}
\int_{-\infty}^{\infty}
S_{xx}(\omega)\,d\omega,\nonumber\\
&\langle p^2\rangle
=
\frac{m^2}{2\pi}
\int_{-\infty}^{\infty}
\omega^2S_{xx}(\omega)\,d\omega .
\end{align}
The effective occupation is then
\begin{equation}
n_{\rm eff}
=
\frac{1}{\hbar\omega_0}
\left(
\frac{\langle p^2\rangle}{2m}
+
\frac{m\omega_0^2\langle x^2\rangle}{2}
\right)
-\frac12 .
\label{eq:neff_def}
\end{equation}
We then optimize $n_{\rm eff}$ over $(\gamma_{\rm fb},\omega_f)$ for fixed $(n_{\rm th},S_{xx}^{\rm imp},\eta)$ subject to Eq.~\eqref{eq:quantum_limit}.

\begin{figure*}[!h]
  \centering
  \includegraphics[width=0.85\textwidth]{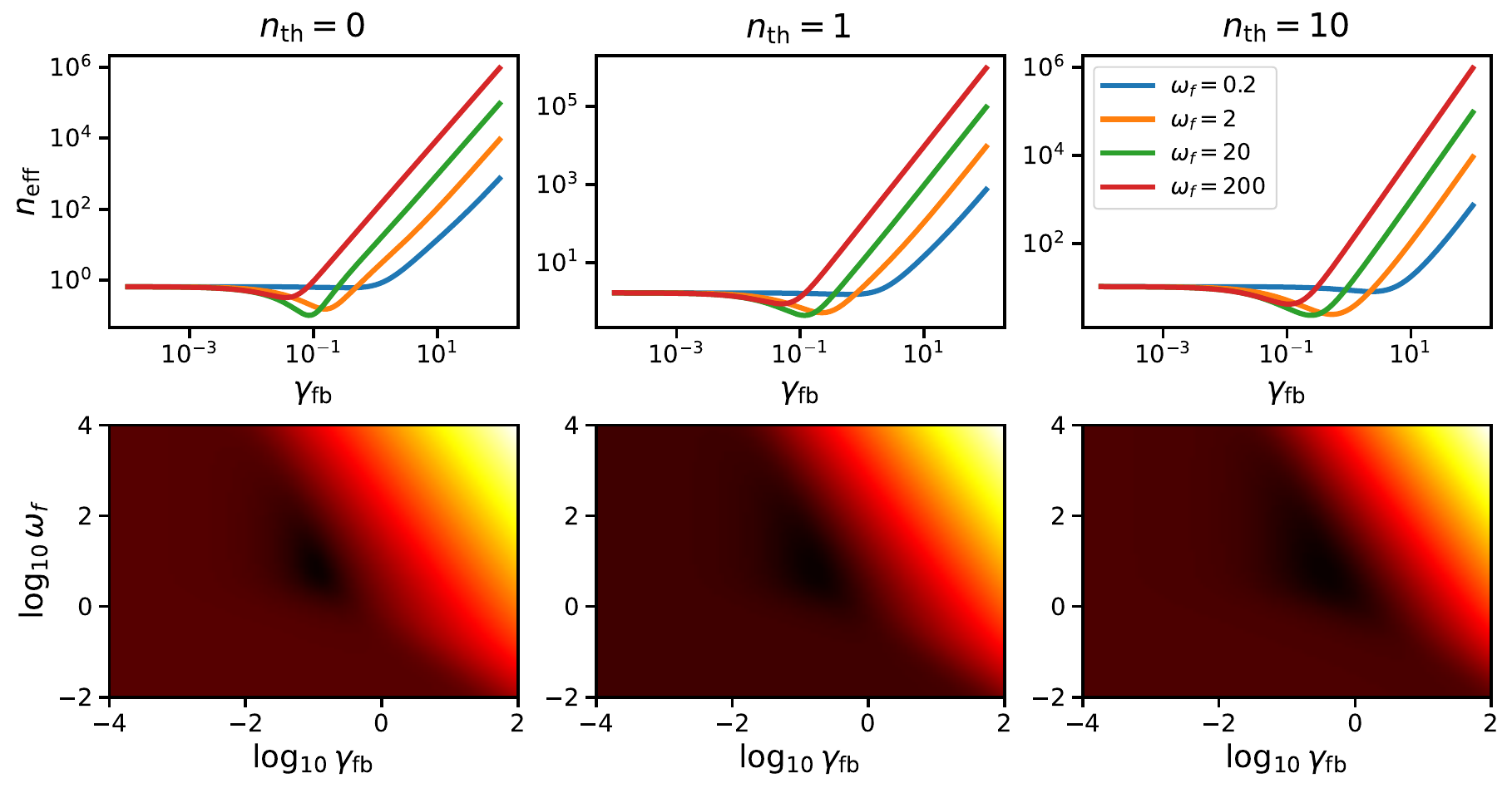}
  \vspace{-1mm}
    \caption{
    Quantum-limited measurement-based band-limited derivative feedback for a harmonic oscillator with thermal forcing set by $n_{\rm th}=0$, $1$, and $10$ from left to right.
    Top row: effective occupation $n_{\rm eff}$ versus feedback damping gain $\gamma_{\rm fb}$ for derivative cutoffs $\omega_f\in\{0.2,2,20,200\}\,\mathrm{rad/s}$.
    Bottom row: optimized landscape $n_{\rm eff}(\gamma_{\rm fb},\omega_f)$ in the $(\gamma_{\rm fb},\omega_f)$ design plane.
    In the bottom row, color encodes $\log_{10}(n_{\rm eff})$ on the same scale in all three panels, with darker colors indicating lower occupation.
    The non-monotonic dependence on $\gamma_{\rm fb}$ and the finite optimal range of $\omega_f$ illustrate the tradeoff between feedback-induced damping and imprecision-noise reinjection.
    }
  \label{fig:quantum_feedback_panels}
\end{figure*}

The loss of monotonic improvement is identified by non-monotonic $n_{\rm eff}(\gamma_{\rm fb})$ and by degradation of performance as $\omega_f\to\infty$ under the white-imprecision idealization: at sufficiently large derivative gain or cutoff, the added imprecision reinjection outweighs the improvement in damping.
These gain- and bandwidth-dependent tradeoffs are summarized in Fig.~\ref{fig:quantum_feedback_panels}.

\subsection{Full-spectrum evaluation}
\label{sec:full_spectrum_bandwidth}

The closed-loop spectrum above captures the tradeoff between feedback-induced damping and measurement-noise reinjection. 
Because derivative feedback weights high frequencies strongly, we additionally evaluate the steady state using the full displacement spectrum and introduce an explicit measurement bandwidth that regularizes the high-frequency behavior.

\paragraph*{Finite measurement bandwidth.}

\begin{figure*}[!h]
  \centering
    \includegraphics[width=0.99\textwidth]{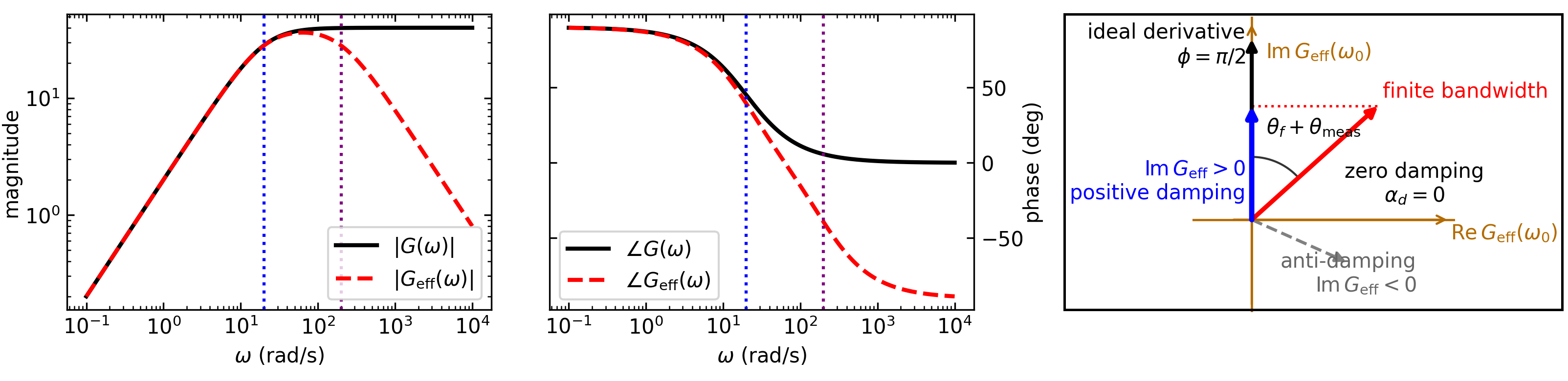}
    \vspace{-1mm}
    \caption{Comparison of the derivative controller transfer function $G(\omega)$ and the effective loop filter $G_{\rm eff}(\omega)=G(\omega)H_{\rm meas}(\omega)$ when the finite bandwidth of the detector and readout chain, $H_{\rm meas}(\omega)=[1+i\omega/\omega_{\rm meas}]^{-1}$, is included. 
    Magnitude (left) and phase (middle) for $G(\omega)$ (black solid line) and $G_{\rm eff}(\omega)$ (red dashed line) with $\omega_f$ (blue dotted line) and $\omega_{\rm meas}$ (purple dotted line). 
    For $\omega\ll \omega_{\rm meas}$, $G_{\rm eff}\approx G$, while for $\omega\gg\omega_{\rm meas}$ the additional roll-off suppresses high-frequency reinjection of imprecision noise and introduces extra phase lag.
    The right panel gives a phasor interpretation of the effective response at $\omega=\omega_0$.
    An ideal derivative lies along the positive imaginary axis, while the finite controller and measurement bandwidths introduce the phase lags $\theta_f=\arctan(\omega_0/\omega_f)$ and $\theta_{\rm meas}=\arctan(\omega_0/\omega_{\rm meas})$, rotating the effective response toward the real axis.
    The imaginary projection of $G_{\rm eff}(\omega_0)$ gives the component responsible for near-resonant damping.
    At the real-axis boundary this component vanishes; further phase rotation changes its sign and produces anti-damping.
    }
  \label{fig:loop_filter_comparison}
\end{figure*}

We model the finite bandwidth of the detector and readout chain by filtering the measurement record through the measurement-bandwidth response $H_{\rm meas}(\omega)$ prior to actuation,
\begin{equation}
y_{\rm fb}(\omega)=H_{\rm meas}(\omega)\,y(\omega),\quad
H_{\rm meas}(\omega)=\frac{1}{1+i\omega/\omega_{\rm meas}},
\label{eq:Hmeas}
\end{equation}
and apply the same band-limited derivative controller as before,
\begin{equation}
F_{\rm fb}(\omega)=-G(\omega)\,y_{\rm fb}(\omega),
\label{eq:Gband-limited_again}
\end{equation}
where $G(\omega)$ is given by Eq.~\eqref{eq:band-limited_derivative}.
This modifies the feedback pathway through the effective loop filter  $G_{\rm eff}(\omega)=G(\omega)H_{\rm meas}(\omega)$, which determines both the closed-loop susceptibility modification and the reinjection of measurement imprecision.
The corresponding force-noise contribution associated with imprecision becomes $|G_{\rm eff}(\omega)|^2 S_{xx}^{\rm imp}(\omega)$.
In the limit $\omega_{\rm meas}\to\infty$, $H_{\rm meas}\to 1$ and one recovers the white-imprecision idealization used above, whereas finite $\omega_{\rm meas}$ provides a physically motivated high-frequency cutoff (Fig.~\ref{fig:loop_filter_comparison}).

\paragraph*{Full-spectrum $n_{\rm eff}$.}

With finite measurement bandwidth, the full-spectrum calculation replaces the feedback filter $G(\omega)$ in Eq.~\eqref{eq:Sxx_closed} by the effective loop filter $G_{\rm eff}(\omega)=G(\omega)H_{\rm meas}(\omega)$:
\begin{align}
S_{xx}(\omega)
=
&\left|
\frac{\chi_m(\omega)}
{1+\chi_m(\omega)G_{\rm eff}(\omega)}
\right|^2
(S_{FF}^{\rm th}
\nonumber\\
&+S_{FF}^{\rm ba}+|G_{\rm eff}(\omega)|^2S_{xx}^{\rm imp}).
\label{eq:Sxx_with_bandwidth}
\end{align}
Here $F_{\rm th}$ and $F_{\rm ba}$ act directly on the mechanical oscillator and do not pass through $H_{\rm meas}$ (cf. Fig.~\ref{fig:feedback_loop_schematic}).
Only the measured record $y(\omega)$, defined in Eq.\,\eqref{eq:classical_meas_record}, passes through this filter before the derivative controller, so its imprecision component is reinjected through $G_{\rm eff}=G H_{\rm meas}$.
The displacement response to each force contribution is shaped by the closed-loop susceptibility in Eq.~\eqref{eq:Sxx_with_bandwidth}, obtained from Eq.~\eqref{eq:classical_chi_eff} with $G$ replaced by $G_{\rm eff}$.

We compute $\langle x^2\rangle$ and $\langle p^2\rangle$ by integrating Eq.~\eqref{eq:Sxx_with_bandwidth} over $\omega$ and extract $n_{\rm eff}$ from Eq.~\eqref{eq:neff_def}.
This full-spectrum evaluation provides the reference calculation against which the high-$Q$ design rule below is compared, and makes explicit how finite $\omega_{\rm meas}$ regularizes the high-frequency contribution of reinjected measurement-imprecision noise.

\begin{figure*}[h]
\centering
\includegraphics[width=0.85\textwidth]{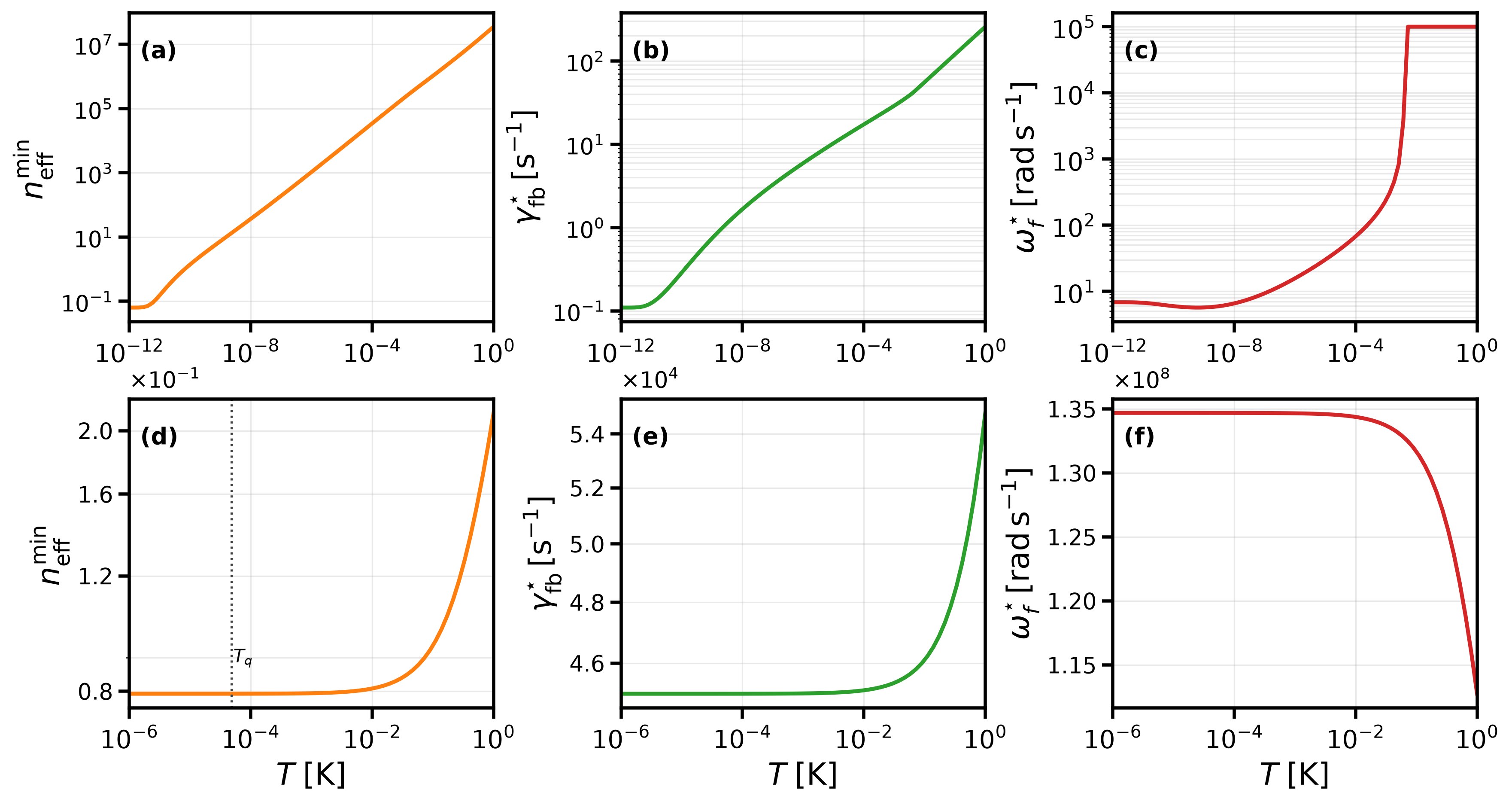}
\vspace{-1mm}
    \caption{
    Full-spectrum optimization of band-limited derivative feedback parameters for a quantum-limited position detector.
    Panels (a)--(c) show the original low-frequency illustrative case, while panels (d)--(f) show a representative MHz-scale case.
    The left column gives the minimum effective occupation $n_{\mathrm{eff}}^{\min}$ obtained from numerical integration of the steady-state variances and optimization over $(\gamma_{\mathrm{fb}},\omega_f)$.
    The middle column gives the optimal feedback-damping gain $\gamma_{\mathrm{fb}}^\star$, and the right column gives the optimal derivative cutoff $\omega_f^\star$.
    For the upper row, $m=1~\mathrm{kg}$, $\omega_0=2~\mathrm{rad/s}$, $\gamma_u=0.05~\mathrm{s^{-1}}$, $\eta=1$, $S_{xx}^{\rm imp}=10^{-34}~\mathrm{m^2\,s}$, and $\omega_{\rm meas}=200~\mathrm{rad/s}$.
    For the lower row, $\omega_0/(2\pi)=1~\mathrm{MHz}$, $m=10^{-12}~\mathrm{kg}$, $Q=10^7$, $\eta=0.8$, $S_{xx}^{\rm imp}=10^{-34}~\mathrm{m^2\,s}$, and $\omega_{\rm meas}/\omega_0=10$.
    In both cases the imprecision--backaction product is taken to saturate Eq.~\eqref{eq:quantum_limit}.
    The vertical dotted line in panel (d) marks $T_q=\hbar\omega_0/k_B\simeq48~\mu\mathrm{K}$ for the MHz-scale example.
    }
\label{fig:quantum_feedback_opt_vs_T}
\end{figure*}

The optimized full-spectrum trends versus temperature are shown in Fig.~\ref{fig:quantum_feedback_opt_vs_T}. 
The upper row retains the low-frequency illustrative example used above, while the lower row gives an additional MHz-scale example using a representative MHz-scale parameter set.
The analytic phase-lag-corrected estimate is derived below.

The MHz-scale example also illustrates the reduction of the thermal occupation at cryogenic temperatures.
For $\omega_0/(2\pi)=1~\mathrm{MHz}$, the exact Bose--Einstein occupation is $n_{\rm th}\simeq1.62$ at $T=100~\mu\mathrm{K}$ and $n_{\rm th}\simeq20.3$ at $T=1~\mathrm{mK}$.
For the representative parameters used in Fig.~\ref{fig:quantum_feedback_opt_vs_T}, the full-spectrum optimization gives a finite optimal controller cutoff: $r_f^\star=\omega_f^\star/\omega_0$ decreases from approximately $21.4$ at $100~\mu\mathrm{K}$ to $17.9$ at $1~\mathrm{K}$, corresponding to a decrease in $\omega_f^\star/(2\pi)$ from approximately $21.4~\mathrm{MHz}$ to $17.9~\mathrm{MHz}$.
Thus, in this example the optimal controller bandwidth is approximately twenty times the mechanical frequency rather than being pushed toward the $\omega_f\rightarrow\infty$ limit.
This parameter set is intended as a representative MHz-scale example rather than a reconstruction of a specific experiment.

\subsection{Phase lag and separation of damping from noise reinjection}
\label{sec:phase_lag_bandwidth}

The finite measurement bandwidth introduced above does more than regularize high-frequency imprecision noise: it also adds phase lag in the feedback path.
As a result, the factor that controls feedback-induced damping is not generally identical to the factor that controls imprecision-noise reinjection. 
This distinction is important near resonance, where derivative feedback is intended to act predominantly in the velocity quadrature.
To make this explicit, evaluate the effective loop filter at $\omega\simeq\omega_0$,
\begin{equation}
G_{\rm eff}(\omega)=
\frac{2m\gamma_{\rm fb} i\omega}
{(1+i\omega/\omega_f)(1+i\omega/\omega_{\rm meas})}.
\label{eq:Geff_phase_lag}
\end{equation}
Defining
\begin{equation}
q_f\equiv\frac{\omega_0}{\omega_f},
\qquad
q_{\rm meas}\equiv\frac{\omega_0}{\omega_{\rm meas}},
\label{eq:q_f_q_meas}
\end{equation}
one finds at resonance
\begin{equation}
G_{\rm eff}(\omega_0)
=
2m\gamma_{\rm fb}\omega_0
\frac{
(q_f+q_{\rm meas})+i(1-q_f q_{\rm meas})
}
{(1+q_f^2)(1+q_{\rm meas}^2)}.
\label{eq:Geff_resonance_decomposition}
\end{equation}
With the sign convention of Eq.~\eqref{eq:classical_chi_m}, the imaginary part of $G_{\rm eff}(\omega_0)$ contributes to viscous damping, while $|G_{\rm eff}(\omega_0)|^2$ controls the strength of imprecision-noise reinjection.
Thus it is useful to define two separate bandwidth factors,
\begin{equation}
\alpha_d
\equiv
\frac{\operatorname{Im}\,G_{\rm eff}(\omega_0)}
{2m\gamma_{\rm fb}\omega_0}
=
\frac{1-q_f q_{\rm meas}}
{(1+q_f^2)(1+q_{\rm meas}^2)},
\label{eq:alpha_d}
\end{equation}
and
\begin{equation}
\alpha_n
\equiv
\frac{|G_{\rm eff}(\omega_0)|^2}
{(2m\gamma_{\rm fb}\omega_0)^2}
=
\frac{1}
{(1+q_f^2)(1+q_{\rm meas}^2)}.
\label{eq:alpha_n}
\end{equation}
Here $\alpha_d$ measures how efficiently the feedback increases the mechanical damping, whereas $\alpha_n$ measures how strongly measurement imprecision is converted into force noise. 
In the ideal broadband limit, $q_f,q_{\rm meas}\ll 1$, both factors approach unity. 
For finite bandwidths, however, they differ: phase lag reduces $\alpha_d$ more strongly than $\alpha_n$.

The condition for derivative feedback to provide positive near-resonant damping is
\begin{equation}
\alpha_d>0
\Longleftrightarrow
q_f q_{\rm meas}<1
\Longleftrightarrow
\omega_f\omega_{\rm meas}>\omega_0^2 .
\label{eq:positive_damping_condition}
\end{equation}
The origin of this condition can also be understood directly from the feedback phase.
At resonance, $\arg\{G_{\rm eff}(\omega_0)\} ={\pi}/{2}-\arctan(q_f)-\arctan(q_{\rm meas}).$
The ideal derivative contributes a phase of $\pi/2$, while the controller and measurement low-pass responses introduce the phase lags $\arctan(q_f)$ and $\arctan(q_{\rm meas})$, respectively.
As illustrated in Fig.~\ref{fig:loop_filter_comparison}, the imaginary component of $G_{\rm eff}(\omega_0)$ is the damping-producing part of the near-resonant feedback response.
Positive damping therefore requires $\operatorname{Im}G_{\rm eff}(\omega_0)>0$.
The boundary occurs when the two filter phase lags together consume the full $\pi/2$ derivative phase, $\arctan(q_f)+\arctan(q_{\rm meas})=\pi/2$.
For positive $q_f$ and $q_{\rm meas}$, this is equivalent to $q_fq_{\rm meas}=1$, or $\omega_f\omega_{\rm meas}=\omega_0^2$.
Beyond this boundary the imaginary component changes sign and the feedback becomes anti-damping near resonance.
Thus finite bandwidth does not merely weaken the feedback loop. 
If the derivative cutoff and measurement bandwidth are both sufficiently low, the effective feedback response is rotated out of the damping quadrature and can become ineffective or even anti-damping near resonance.

With this separation, the high-$Q$ estimate becomes
\begin{equation}
n_{\rm eff}+\frac12
\approx
\frac{A+B_n\gamma_{\rm fb}^2}
{\gamma_u+\alpha_d\gamma_{\rm fb}},
\label{eq:neff_highQ_phase_lag}
\end{equation}
where
\begin{equation}
A=
\gamma_u\left(n_{\rm th}+\frac12\right)
+\frac{S_{FF}^{\rm ba}}{4m\hbar\omega_0},
\quad
B_n=
\frac{m\omega_0\alpha_n S_{xx}^{\rm imp}}{\hbar}.
\label{eq:AB_phase_lag_defs}
\end{equation}
For fixed $(\omega_f,\omega_{\rm meas})$ and $\alpha_d>0$, minimizing Eq.~\eqref{eq:neff_highQ_phase_lag} gives
\begin{equation}
\gamma_{\rm fb}^\star
=\frac{-\gamma_u+
\sqrt{\gamma_u^2+\alpha_d^2 A/B_n}}
{\alpha_d}.
\label{eq:gamma_star_phase_lag}
\end{equation}
For $\alpha_d\le 0$, the near-resonant feedback does not provide positive viscous damping, and the high-$Q$ optimum in Eq.~\eqref{eq:gamma_star_phase_lag} is not physically applicable.
A derivation of this near-resonant estimate is given in Appendix~\ref{app:highQ_reduction}.

\paragraph*{Broadband cold-damping limit.}

As a consistency check, consider the broadband limit $\omega_f,\omega_{\rm meas}\gg\omega_0$, for which $q_f,q_{\rm meas}\rightarrow0$ and hence $\alpha_d,\alpha_n\rightarrow1$.
Equation~\eqref{eq:neff_highQ_phase_lag} then reduces to
\begin{equation}
n_{\rm eff}+\frac12
\approx
\frac{
\gamma_u\left(n_{\rm th}+\frac12\right)
+\dfrac{S_{FF}^{\rm ba}}{4m\hbar\omega_0}
+\dfrac{m\omega_0S_{xx}^{\rm imp}}{\hbar}\gamma_{\rm fb}^2
}
{\gamma_u+\gamma_{\rm fb}} .
\label{eq:neff_broadband_limit}
\end{equation}
This has the standard cold-damping structure: the gain-independent thermal and backaction contributions are reduced by the increased total damping, while measurement imprecision is reinjected through a term quadratic in the feedback gain \cite{CourtyHeidmannPinard2001,GenesPRA2008}.
In the classical, noiseless-readout limit $S_{FF}^{\rm ba},S_{xx}^{\rm imp}\rightarrow0$ and $k_BT\gg\hbar\omega_0$, Eq.~\eqref{eq:neff_broadband_limit} gives, to leading order in $n_{\rm th}\gg1$, $n_{\rm eff}\simeq n_{\rm th}/(1+\gamma_{\rm fb}/\gamma_u)$, equivalent to the familiar cold-damping result for the effective mode temperature, $T_{\rm eff}\simeq T/(1+\gamma_{\rm fb}/\gamma_u)$\cite{CohadonPRL1999,CourtyHeidmannPinard2001,GenesPRA2008}.
The finite-bandwidth result above therefore extends this standard limit by allowing the damping efficiency and imprecision-noise reinjection to acquire the distinct factors $\alpha_d$ and $\alpha_n$.

\paragraph*{Validation against the full-spectrum result.}

\begin{figure*}[!t]
  \centering
  \includegraphics[width=0.95\textwidth]{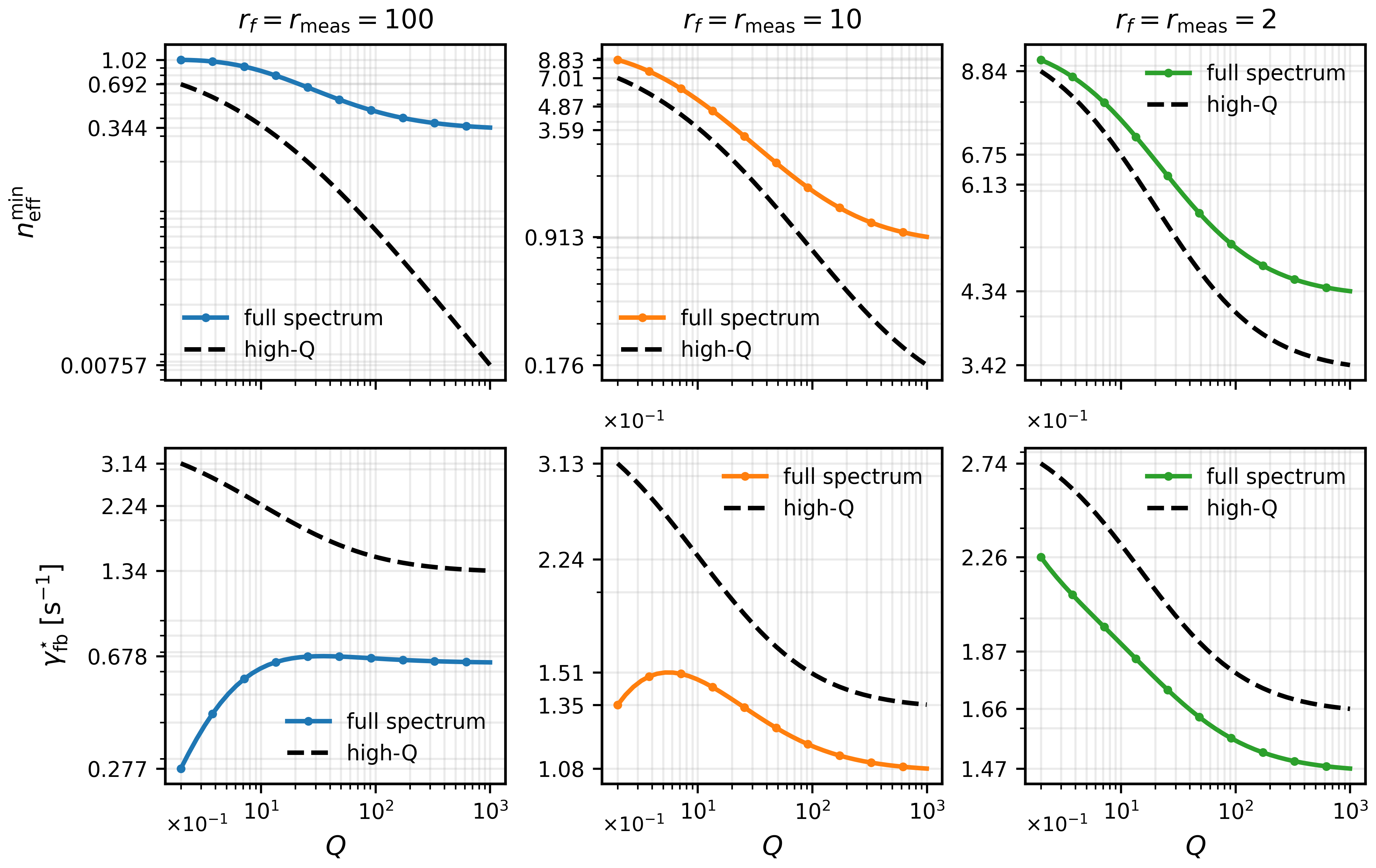}
  \vspace{-1mm}
    \caption{
    Comparison of the high-$Q$ approximation with the full-spectrum calculation at fixed derivative and measurement bandwidths.
    The columns correspond to $r_f=r_{\rm meas}=100$, $10$, and $2$, with $r_f=\omega_f/\omega_0$ and $r_{\rm meas}=\omega_{\rm meas}/\omega_0$.
    The remaining parameters are fixed at $n_{\rm th}=1$, $\eta=1$, $S_{xx}^{\rm imp}=10^{-34}\,\mathrm{m^2\,s}$, $\omega_0=2\,\mathrm{rad/s}$, and $m=1\,\mathrm{kg}$.
    Top row: optimized effective occupation $n_{\rm eff}^{\min}$ as a function of the mechanical quality factor $Q=\omega_0/(2\gamma_u)$.
    Bottom row: corresponding optimal feedback gain $\gamma_{\rm fb}^{\star}$.
    Solid curves show the full-spectrum calculation and dashed curves the high-$Q$ approximation.
    The comparison shows that high mechanical $Q$ alone is not sufficient for quantitative agreement: broad controller and measurement bandwidths retain off-resonant imprecision-noise contributions in the full-spectrum calculation that are not retained in the near-resonant approximation, whereas stronger filtering improves the agreement.
    }
    \label{fig:highQ_full_validation}
\end{figure*}

To assess the quantitative range of the near-resonant approximation, we compare Eqs.~\eqref{eq:neff_highQ_phase_lag} and \eqref{eq:gamma_star_phase_lag} directly with the full-spectrum calculation at fixed controller and measurement bandwidths.
For this comparison we use the bandwidth ratios
\begin{equation}
r_f\equiv \frac{\omega_f}{\omega_0},
\qquad
r_{\rm meas}\equiv \frac{\omega_{\rm meas}}{\omega_0}.
\label{eq:dimensionless_ratios}
\end{equation}
Fig.~\ref{fig:highQ_full_validation} shows the optimized occupation and feedback gain as functions of the mechanical quality factor $Q=\omega_0/(2\gamma_u)$ for three representative choices $r_f=r_{\rm meas}\in\{100,10,2\}$.
For the more strongly filtered case, $r_f=r_{\rm meas}=2$, the high-$Q$ estimate remains comparatively close to the full-spectrum result, whereas the discrepancy becomes substantially larger as the controller and measurement bandwidths are increased.
In particular, increasing $Q$ alone does not guarantee convergence of the narrowband estimate to the full-spectrum result.
For broad feedback and measurement bandwidths, the remaining discrepancy is consistent with off-resonant imprecision-noise contributions retained by the full-spectrum integral but neglected in the near-resonant approximation.
The comparison therefore shows that the accuracy of the compact design rule is controlled not only by the mechanical quality factor but also by the bandwidth ratios $r_f$ and $r_{\rm meas}$.
The full-spectrum curves are evaluated using the algebraic moments and stable-gain optimization described in Appendix~\ref{app:full_spectrum_moments}.

Equations~\eqref{eq:alpha_d}--\eqref{eq:gamma_star_phase_lag} show that the optimal controller is determined by two competing bandwidth effects: finite bandwidth suppresses imprecision reinjection through $\alpha_n$, but it also reduces useful damping through $\alpha_d$. 
This damping--noise separation is the main finite-bandwidth correction and is used below to construct the dimensionless design map.

Table~\ref{tab:symbol_summary} in Appendix~\ref{app:notation} summarizes the notation used in Secs.~\ref{sec:phase_lag_bandwidth}--\ref{sec:nn_surrogate}.

\subsection{Dimensionless controller-design map}
\label{sec:dimensionless_map}

To present the design problem in a form that is portable across oscillator frequencies and experimental platforms, we use the bandwidth ratios $r_f$ and $r_{\rm meas}$ introduced in Sec.~\ref{sec:phase_lag_bandwidth}.
Equivalently, the inverse ratios used above are $q_f=1/r_f$ and $q_{\rm meas}=1/r_{\rm meas}$.

In the high-$Q$ approximation, finite controller and measurement bandwidths enter through two distinct near-resonant factors. 
The useful damping is controlled by
\begin{equation}
\alpha_d(r_f,r_{\rm meas})
=\frac{1-1/(r_f r_{\rm meas})}
{\left[1+1/r_f^2\right]\left[1+1/r_{\rm meas}^2\right]},
\label{eq:alpha_d_dimensionless}
\end{equation}
whereas the imprecision-noise reinjection is controlled by
\begin{equation}
\alpha_n(r_f,r_{\rm meas})
=
\frac{1}
{\left[1+1/r_f^2\right]\left[1+1/r_{\rm meas}^2\right]} .
\label{eq:alpha_n_dimensionless}
\end{equation}
Thus finite bandwidth has two competing effects: it suppresses reinjected imprecision noise through $\alpha_n$, but it also rotates the effective feedback response away from the velocity quadrature and reduces the useful damping through $\alpha_d$.

The positive-damping condition becomes
\begin{equation}
\alpha_d>0
\quad\Longleftrightarrow\quad
r_f r_{\rm meas}>1 .
\label{eq:positive_damping_dimensionless}
\end{equation}
This boundary separates the design plane into a positive-damping region, where derivative feedback increases the near-resonant damping, and a phase-lag-dominated region, where the feedback no longer provides positive near-resonant damping.

\begin{figure*}[!h]
  \centering \includegraphics[width=\textwidth]{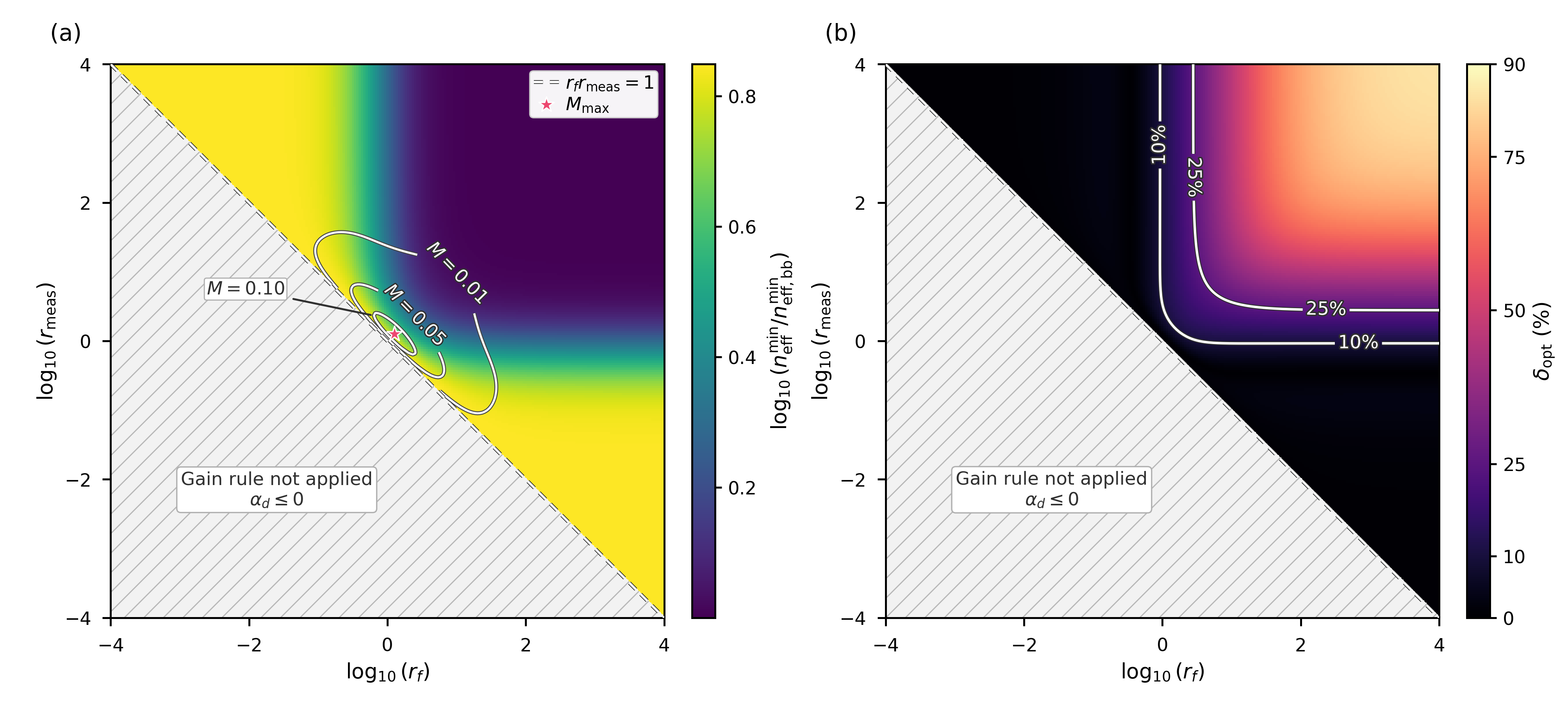}
    \caption{
    Dimensionless controller-design map for band-limited derivative feedback with finite measurement bandwidth.
    (a) Gain-optimized high-$Q$ occupation relative to its broadband benchmark ($\alpha_d=\alpha_n=1$), $\log_{10}[n_{\rm eff}^{\min}/n_{\rm eff,bb}^{\min}]$.
    Solid contours show $M=g^\star/g_{\rm RH}$, where $g^\star=\gamma_{\rm fb}^\star/\omega_0$ and $g_{\rm RH}$ is the full closed-loop stability limit.
    The star marks the grid maximum, $M_{\max}\simeq0.124$.
    (b) Relative discrepancy $\delta_{\rm opt}=|n_{\rm H}-n_{\rm F}|/n_{\rm F}$ in percent, where $n_{\rm H}\equiv n_{\rm eff}^{\min}$ and $n_{\rm F}$ are the high-$Q$ and full-spectrum minima at fixed bandwidths.
    The latter is optimized over $0\le g<g_{\rm RH}$.
    White contours mark $10\%$ and $25\%$ error, not instability.
    The dashed boundary $r_f r_{\rm meas}=1$ marks the near-resonant positive-damping condition.
    Hatching denotes where the analytical gain rule is not applied, not closed-loop instability.
    Parameters: $Q=20$, $n_{\rm th}=1$, $\eta=1$, $m=1~\mathrm{kg}$,    $\omega_0=2~\mathrm{rad/s}$, and $S_{xx}^{\rm imp}=10^{-34}~\mathrm{m^2\,s}$.
    }
    \label{fig:dimensionless_map}
\end{figure*}

Using the high-$Q$ estimate in Eq.~\eqref{eq:neff_highQ_phase_lag}, with $\alpha_d$ and $\alpha_n$ evaluated from Eqs.~\eqref{eq:alpha_d_dimensionless} and \eqref{eq:alpha_n_dimensionless}, we minimize over $\gamma_{\rm fb}$ at fixed $(r_f,r_{\rm meas})$.
We evaluate the positive-gain solution in Eq.~\eqref{eq:gamma_star_phase_lag} only for $\alpha_d>0$, where $\gamma_{\rm fb}^\star>0$ and the near-resonant total amplitude-damping rate $\gamma_u+\alpha_d\gamma_{\rm fb}^\star$ is positive.
The region $\alpha_d\le0$, where this gain rule is not applied, is hatched in Fig.~\ref{fig:dimensionless_map}.
Full closed-loop stability is checked separately below.
The map therefore summarizes not only the loss of loop strength at small bandwidth, but also the separation between useful damping and imprecision-noise reinjection.
Within this high-$Q$ approximation, the optimized occupation decreases as either bandwidth increases at fixed remaining parameters.
Departures from this near-resonant prediction require a full-spectrum assessment of off-resonant noise contributions.

\paragraph*{Closed-loop stability.}

The condition $r_f r_{\rm meas}>1$ determines the sign of the near-resonant feedback-damping contribution, but does not by itself guarantee closed-loop stability at arbitrary gain.
For $g=\gamma_{\rm fb}/\omega_0$ and $\epsilon=\gamma_u/\omega_0>0$, the full closed-loop model is asymptotically stable when $0\le g<g_{\rm RH}$.
The characteristic polynomial and the gain limit $g_{\rm RH}(r_f,r_{\rm meas},\epsilon)$ are derived in Appendix~\ref{app:closed_loop_stability}, with the limit given by Eq.~\eqref{eq:app_g_RH}.

For Fig.~\ref{fig:dimensionless_map}, $\epsilon=0.025$ and $Q=20$.
The calculation uses a $241\times241$ grid spanning $-4\le\log_{10}r_f,\log_{10}r_{\rm meas}\le4$.
Here $g^\star=\gamma_{\rm fb}^\star/\omega_0$ denotes the normalized gain predicted by the high-$Q$ rule in Eq.~\eqref{eq:gamma_star_phase_lag}.
All $28\,920$ evaluated high-$Q$ optimal gains satisfy $g^\star<g_{\rm RH}$.
Direct calculation of the four characteristic roots confirms negative real parts at every evaluated point.
Defining $M=g^\star/g_{\rm RH}$, the largest ratio on the evaluated grid is $M=0.12398$, at $r_f=r_{\rm meas}\simeq1.259$.
The smallest multiplicative increase in gain required to reach the instability threshold, with the other parameters fixed, is therefore approximately $8.07$.
The solid contours in Fig.~\ref{fig:dimensionless_map}\,(a) show $M$.
No $M=1$ contour occurs on the evaluated grid.
This stability check does not establish the quantitative accuracy of the high-$Q$ occupation estimate.

\paragraph*{Reliability of the high-$Q$ design map.}

We compare the high-$Q$ minimum shown in Fig.~\ref{fig:dimensionless_map}\,(a), denoted here by $n_{\rm H}\equiv n_{\rm eff}^{\min}$, with the full-spectrum minimum at the same fixed bandwidths,
\begin{equation}
n_{\rm F}(r_f,r_{\rm meas})=
\min_{0\le g<g_{\rm RH}}n_{\rm eff}^{\rm full}(g;r_f,r_{\rm meas}).
\label{eq:map_full_minimum}
\end{equation}
At each fixed pair $(r_f,r_{\rm meas})$, we optimize only the feedback gain $g=\gamma_{\rm fb}/\omega_0$.
The oscillator, bath, and detector parameters are the same as those used for the high-$Q$ map in Fig.~\ref{fig:dimensionless_map}\,(a).
The full-frequency moments are evaluated algebraically for the same uncorrelated, constant-noise model, without the high-$Q$ approximation (Appendix~\ref{app:full_spectrum_moments}).
All stationary candidates in the stable gain interval and the zero-gain endpoint are compared.
We independently check the occupations at both selected gains using covariance calculations at $84$ bandwidth pairs and frequency quadrature at $42$ pairs.
The discrepancy is quantified by
\begin{equation}
\delta_{\rm opt}=\frac{|n_{\rm H}-n_{\rm F}|}{n_{\rm F}}.
\label{eq:map_relative_occupation_error}
\end{equation}
Figure~\ref{fig:dimensionless_map}\,(b) displays this discrepancy in percent.
Of the $28\,920$ evaluated points, $14\,185$ ($49.05\%$) satisfy $\delta_{\rm opt}\le0.10$, and $17\,687$ ($61.16\%$) satisfy $\delta_{\rm opt}\le0.25$.
These are fractions of the sampled logarithmic grid at $Q=20$, not universal bandwidth criteria.
The largest discrepancy is $85.09\%$ at $r_f=r_{\rm meas}=10^4$, where $n_{\rm H}\simeq0.235$ and $n_{\rm F}\simeq1.576$.
The broad, approximately equal-bandwidth region therefore has the largest occupation errors despite the stability of the analytical gains.
Small discrepancies in strongly asymmetric regions can instead occur when both occupations remain close to the zero-feedback value $n_0=n_{\rm eff}^{\rm full}(g=0)$, with measurement backaction retained.
The discrepancy between optimized occupations is distinct from the performance of the analytical controller.
We also evaluate the full-spectrum occupation at the gain predicted by the high-$Q$ rule, $n_{\rm F|H}=n_{\rm eff}^{\rm full}(g^\star)$.
At $r_f=r_{\rm meas}=10^4$, this gives $n_{\rm F|H}\simeq178.25$, compared with $n_{\rm F}\simeq1.576$ and the zero-feedback occupation $n_0\simeq1.659$.
Thus, stable analytical gains need not be nearly optimal for the full-frequency model.

The bandwidth scales relevant to the design map span a broad range in current mechanical platforms.
For example, a sideband-resolved optomechanical-crystal cavity reported by Ren \emph{et al.} has $\omega_0/2\pi=10.02~\mathrm{GHz}$ and an optical linewidth $\kappa/2\pi=1.19~\mathrm{GHz}$ \cite{RenNatCommun2020}, giving a linewidth-to-mechanical-frequency ratio $\kappa/\omega_0\simeq0.12$.
Although this experiment does not implement the direct measurement-based feedback considered here, it illustrates that optomechanical readout dynamics can occur on a frequency scale below the mechanical resonance.
In contrast, the measurement-based feedback experiment of Rossi \emph{et al.} used a membrane mode at $\omega_0/2\pi=1.139~\mathrm{MHz}$ and a probe-cavity linewidth $\kappa/2\pi=15.9~\mathrm{MHz}$ \cite{RossiNature2018}, corresponding to $\kappa/\omega_0\simeq14$, while the feedback filter and loop delay were engineered on frequency and time scales relevant to the mechanical resonance.
Low-frequency macroscopic systems can likewise operate with measurement and control scales of order $\omega_0$: Agafonova \emph{et al.} used an $18~\mathrm{Hz}$ torsional mode within a low-noise measurement band extending approximately from $8$ to $28~\mathrm{Hz}$ \cite{AgafonovaCommunPhys2026}, and Whittle \emph{et al.} controlled a $10~\mathrm{kg}$ interferometric test mass trapped near $148~\mathrm{Hz}$ with feedback dynamics shaped over approximately $100$--$200~\mathrm{Hz}$ \cite{WhittleScience2021}.
These examples should not be interpreted as exact coordinates in the $(r_f,r_{\rm meas})$ plane, since practical measurement and feedback paths generally contain cavity dynamics, band-pass filtering, additional poles and zeros, actuator dynamics, and explicit delays rather than the two single-pole responses assumed here.
They nevertheless show that experimentally relevant readout and control bandwidths can range from below the mechanical frequency to much larger than it, with several platforms having dynamical scales comparable to $\omega_0$.
These examples show that the finite-bandwidth regime represented by the design map is not restricted to an artificial parameter limit, and the phase-lag constraint identified here can become relevant whenever either the measurement or feedback response is not broadband compared with the mechanical resonance.
Within the simplified single-pole model, systems with measurement and controller response scales comparable to $\omega_0$ are the most likely to approach the $r_f r_{\rm meas}=1$ boundary, whereas broadband measurement and control place the system well inside the positive-damping region.
The design map, like the preceding quantitative results, assumes the uncorrelated detector model; Appendix~\ref{app:quantum_correlations} shows how imprecision--backaction correlations can modify the optimum.
Specifically, Figs.~\ref{fig:quantum_feedback_panels}, \ref{fig:quantum_feedback_opt_vs_T}, \ref{fig:highQ_full_validation}, and \ref{fig:dimensionless_map} assume vanishing imprecision--backaction cross-correlation, $\rho=0$, using the notation introduced in Appendix~\ref{app:quantum_correlations}.

\subsection{Neural-network surrogate}
\label{sec:nn_surrogate}

The analytic results above identify the physical origin of the feedback limitation through the separation of useful damping and imprecision-noise reinjection. 
For the surrogate dataset used here, the full-spectrum targets were generated by numerical integration of Eq.~\eqref{eq:Sxx_with_bandwidth} and minimization over $(\gamma_{\rm fb},\omega_f)$ for each detector and bath parameter set.
We therefore train a small neural-network surrogate for the full-spectrum optimizer. 
Fully connected feed-forward networks are standard nonlinear function approximators \cite{Cybenko1989,Hornik1989,Bishop1995}, and here they are used only as an interpolation tool for numerically optimized design data. 

\begin{figure*}[!h]
  \centering
  \includegraphics[width=0.9\textwidth]{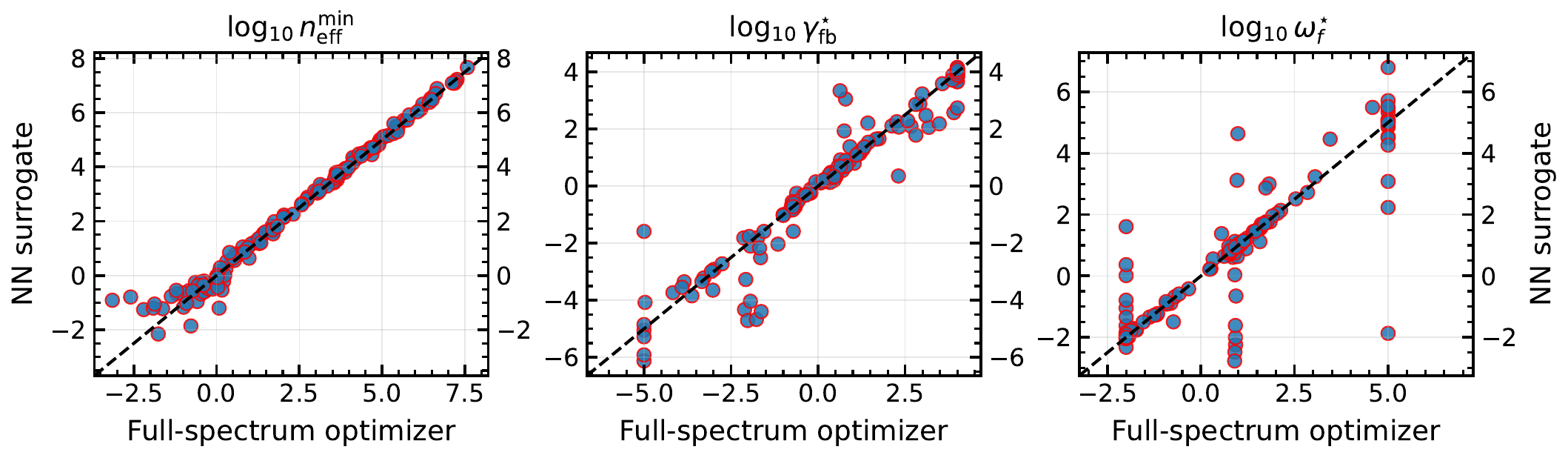}
    \caption{
    Neural-network surrogate for first-pass full-spectrum design estimates.
    The three panels compare neural-network predictions with held-out full-spectrum optimization results for $n_{\rm eff}^{\min}$, $\gamma_{\rm fb}^{\star}$, and $\omega_f^{\star}$.
    The mean absolute errors reported in the text are computed exclusively on the held-out test set.
    The dashed line indicates perfect agreement.
    The surrogate captures the optimized occupation most accurately, gives a useful first-pass estimate of the feedback gain, and is less precise for the cutoff $\omega_f^{\star}$, consistent with the shallow dependence of the full-spectrum objective on $\omega_f$ in parts of the design space.
    }
  \label{fig:nn_surrogate_parity}
\end{figure*}

The network input is
\begin{equation} 
\mathbf{z}
=
\left(
\log_{10}\!\left(n_{\rm th}+\frac{1}{2}\right),
\log_{10} S_{xx}^{\rm imp},
\eta,
\log_{10}\!\left(\frac{\omega_{\rm meas}}{\omega_0}\right)
\right),
\label{eq:nn_input_vector}
\end{equation}
and the target output is
\begin{equation}
\mathbf{o}
=
\left(
\log_{10} n_{\rm eff}^{\min},
\log_{10} \gamma_{\rm fb}^{\star},
\log_{10} \omega_f^{\star}
\right).
\label{eq:nn_output_vector} 
\end{equation}
These targets are obtained by numerical integration and optimization, rather than from the high-$Q$ approximation.
Thus, the neural network does not replace the analytic phase-lag criterion; it provides a rapid first-pass estimate of the full-spectrum optimum over a multidimensional parameter space. 
We train separate multilayer-perceptron regressors for the three output quantities using standardized inputs and logarithmic targets, implemented with the \texttt{scikit-learn} machine-learning library \cite{Pedregosa2011}. 
The dataset contains 700 full-spectrum optimization samples and is split into 560 training samples and 140 held-out test samples (80\%/20\%), using a shuffled split with \texttt{random\_state}=123.
During cutoff-model selection, the training portion was further divided into 448 model-selection training samples and 112 validation samples using \texttt{random\_state}=456. 
After selecting the architecture using validation mean absolute error (MAE), the final cutoff regressor used for Fig.~\ref{fig:nn_surrogate_parity} was freshly initialized and retrained on all 560 outer-training samples.

The input standardization is fitted using the 560-sample training set only, with a separate \texttt{StandardScaler} for the inputs of each regressor and a separate \texttt{StandardScaler} for each logarithmic target.
The target standardization is inverted before evaluation, so that the reported errors are evaluated in $\log_{10}$ units.
Each regressor uses three hidden layers containing 128, 128, and 64 neurons, respectively.
The occupation and gain regressors use ReLU activation and the Adam optimizer, whereas the cutoff regressor uses hyperbolic-tangent activation and the L-BFGS optimizer.
The relevant training parameters are the L2 regularization strength $\alpha=10^{-5}$ and a tolerance of $10^{-4}$ for all three regressors.
The Adam models use a constant learning-rate schedule with an initial learning rate of $5\times10^{-4}$, automatic batch size, a maximum of $12\,000$ iterations, early stopping with a validation fraction of $0.15$ and $400$ no-improvement iterations, and random states 700 and 701.
The L-BFGS cutoff model uses a maximum of $20\,000$ iterations, a maximum of $15\,000$ function evaluations, and random state 999.
The reported prediction errors below are calculated on the held-out test set only.

Figure~\ref{fig:nn_surrogate_parity} compares the neural-network predictions with held-out full-spectrum optimization results. 
The surrogate predicts $n_{\rm eff}^{\min}$ with a mean absolute error of $0.194$ in $\log_{10}$ units, corresponding to a typical multiplicative factor of $1.56$. 
It predicts $\gamma_{\rm fb}^{\star}$ with a mean absolute error of $0.371$ in $\log_{10}$ units, corresponding to a factor of $2.35$. 
The cutoff $\omega_f^{\star}$ is recovered less accurately, with a mean absolute error of $0.462$ in $\log_{10}$ units, corresponding to a factor of $2.89$. 
This reduced accuracy is consistent with the shallow dependence of the full-spectrum objective on $\omega_f$ in parts of the design space and with optima that lie near the imposed search bounds. 

The larger errors in $\gamma_{\rm fb}^{\star}$ and especially $\omega_f^{\star}$ imply that the surrogate is not intended to determine final controller parameters without further numerical refinement.
A multiplicative uncertainty of order $2$--$3$ is nevertheless useful for rapidly identifying the relevant region of parameter space and providing an initial estimate for subsequent full-spectrum optimization.
Accordingly, the neural network is used here as a first-pass estimator, whereas the full-spectrum calculation remains the reference method for final parameter refinement.

The training data, trained model files containing the learned network weights, preprocessing information, and training and reproduction code used to reproduce the neural-network results and Fig.~\ref{fig:nn_surrogate_parity} are publicly available at \textbf{\cite{VujnovicNN2026}}.

\section{Conclusion}

We analyzed measurement-based derivative feedback of a harmonic oscillator in a transfer-function formulation that makes explicit both susceptibility shaping and measurement-noise reinjection. 
For the band-limited derivative controller, the closed-loop displacement spectrum separates into external forcing, measurement backaction in the quantum extension, and an imprecision-reinjection term proportional to the squared loop-filter magnitude. 
Without an additional measurement-bandwidth filter this term scales as $|G(\omega)|^2S_{xx}^{\rm imp}$, while with finite measurement bandwidth it is governed by $|G_{\rm eff}(\omega)|^2S_{xx}^{\rm imp}$. 
This structure exposes a generic design tradeoff: increasing derivative gain broadens the resonance through enhanced effective damping, but also reinjects measurement imprecision, leading to non-monotonic performance versus gain.

The individual ingredients of this problem---measurement-based cold damping, quantum-limited imprecision and backaction, and finite controller and measurement bandwidths---are established in the literature.
The specific result of the present work is the explicit separation of the effects of finite controller and measurement bandwidths into a near-resonant damping factor $\alpha_d$ and an imprecision-reinjection factor $\alpha_n$.
This separation shows that reducing high-frequency noise and preserving useful damping are not equivalent requirements.
It leads directly to the condition for positive near-resonant feedback damping $\omega_f\omega_{\rm meas}>\omega_0^2$ and to the corresponding high-$Q$ optimal-gain rule and dimensionless controller-design map.
The full closed-loop stability analysis confirms that all high-$Q$ gains evaluated in the dimensionless controller-design map are asymptotically stable for the stated parameters.
For the same parameter set, a separate full-spectrum comparison identifies the regions of the evaluated bandwidth grid where the predicted minimum occupation agrees with the full-spectrum minimum to within $10\%$ and $25\%$.
The largest discrepancies occur for broad, approximately equal bandwidths, where stable analytical gains can nevertheless give substantially suboptimal occupations.
Thus, the near-resonant damping-sign condition, full closed-loop stability, and quantitative reliability of the high-$Q$ approximation must be assessed separately.

Finally, we trained a neural-network surrogate on full-spectrum optimization data to provide rapid first-pass design estimates over detector and bath parameters. 
The surrogate is useful for interpolation and initial parameter selection, but it does not replace the analytic phase-lag criterion or the full-spectrum optimizer, which remains the reference calculation for final refinement. 
As illustrated by the additional MHz-scale example, the quantum-limited regime discussed here is more readily accessed for higher-$\omega_0$ oscillators and/or cryogenic temperatures through the mapping $n_{\rm th}(T)=[\exp(\hbar\omega_0/k_B T)-1]^{-1}$.

Natural extensions include replacing the fixed derivative controller with optimal or adaptive feedback strategies and incorporating nonclassical measurement resources such as squeezed measurements.
A complementary direction is to compare the finite-bandwidth limitations identified here for measurement-based feedback with coherent quantum-feedback architectures, in which the control loop does not rely on an explicit measurement record.

\backmatter

\section*{Declarations}

\bmhead{Funding}
No funding was received for this work.

\bmhead{Competing interests}
The author declares no competing interests.

\bmhead{Ethics approval and consent to participate}
Not applicable.

\bmhead{Consent for publication}
Not applicable.

\bmhead{Data availability}
The dataset used to train and test the neural-network surrogate is publicly available at \cite{VujnovicNN2026}.

\bmhead{Code availability}
The neural-network training and reproduction code, together with the trained model files, is publicly available at \cite{VujnovicNN2026}.

\bmhead{Author contributions}
The author conceived the study, developed the model, performed the analytical and numerical calculations, prepared the figures, and wrote the manuscript.

\begin{appendices}
\renewcommand*{\theHequation}{\thesection.\arabic{equation}}
\section{Loop conventions}
\label{app:supp_classical_overview}
\setcounter{equation}{0}

This appendix fixes the classical feedback conventions used in the main text. 
We use a direct-output topology in which a measured displacement record is converted into an actuation force and applied back to the same oscillator mode.

Here we work with transfer functions in the Laplace domain, retaining $s$ as a general complex variable.
The frequency responses used in the main text are obtained by setting $s=i\omega$.
Uppercase signal symbols denote the Laplace transforms of the corresponding time-domain signals.
The linear controller is denoted by $C$, with transfer function $C(s)$.

\subsection{Direct-output feedback topology}
\label{app:topology}

The open-loop oscillator plant is
\begin{equation}
H_0(s)=\frac{1}{m\left(s^2+2\gamma_{u} s+\omega_0^2\right)} .
\label{eq:app_H0}
\end{equation}
Feedback is applied as an additive force at the plant input:
\begin{equation}
f_{\rm tot}(t)=f(t)+u(t).
\label{eq:app_total_force}
\end{equation}
The control force is synthesized from the measured displacement,
\begin{equation}
u(t)=-C\{x_m(t)\}.
\label{eq:app_control_force_time}
\end{equation}
In the Laplace domain this becomes
\begin{equation}
U(s)=-C(s)X_m(s).
\label{eq:app_control_force_laplace}
\end{equation}

For a noise-free measurement, $X_m(s)=X(s)$, the closed-loop response satisfies
\begin{equation}
X(s)=H_0(s)\left[F(s)-C(s)X(s)\right].
\label{eq:app_closed_loop_intermediate}
\end{equation}
Therefore,
\begin{equation}
\frac{X(s)}{F(s)}
=
\frac{H_0(s)}{1+H_0(s)C(s)} .
\label{eq:app_closed_loop_force_response}
\end{equation}
The loop gain is
\begin{equation}
L(s)=H_0(s)C(s),
\label{eq:app_loop_gain}
\end{equation}
and the closed-loop poles are determined by
\begin{equation}
1+L(s)=0.
\label{eq:app_characteristic_equation}
\end{equation}

\subsection{Measurement-noise injection}
\label{app:meas_noise_injection}

Let the measured displacement be
\begin{equation}
x_m(t)=x(t)+n(t),
\label{eq:app_measured_displacement}
\end{equation}
where $n(t)$ is additive measurement noise. 
In the Laplace domain,
\begin{equation}
X_m(s)=X(s)+N(s).
\label{eq:app_measured_displacement_laplace}
\end{equation}
Using $U(s)=-C(s)X_m(s)$, the closed-loop displacement becomes
\begin{equation}
X(s)
=
\frac{H_0(s)}{1+H_0(s)C(s)}F(s)
-
\frac{H_0(s)C(s)}{1+H_0(s)C(s)}N(s).
\label{eq:app_noise_closed_loop}
\end{equation}
With the definitions of the sensitivity function
\begin{equation}
S(s)=\frac{1}{1+L(s)},
\label{eq:app_sensitivity}
\end{equation}
and the complementary sensitivity function
\begin{equation}
T(s)=\frac{L(s)}{1+L(s)},
\label{eq:app_complementary_sensitivity}
\end{equation}
Eq.~\eqref{eq:app_noise_closed_loop} can be written as
\begin{equation}
X(s)=S(s)H_0(s)F(s)-T(s)N(s).
\label{eq:app_sensitivity_noise}
\end{equation}
Thus feedback suppresses external force disturbances through $S(s)$ but reinjects measurement noise through $T(s)$.

\subsection{Filtered derivative controller}
\label{app:filtered_derivative}

An ideal derivative controller is
\begin{equation}
C_D(s)=K_Ds.
\label{eq:app_ideal_derivative}
\end{equation}
Here $K_D$ is the derivative-gain coefficient, corresponding to $2m\gamma_{\rm fb}$ in the main text.
It increases the effective damping of the oscillator, but its gain grows without bound at high frequency. 
A standard regularized form is the filtered derivative controller
\begin{equation}
C_D(s)=\frac{K_Ds}{1+s/\omega_f}.
\label{eq:app_filtered_derivative}
\end{equation}
For $\omega\ll\omega_f$, this behaves as an ordinary derivative.
For $\omega\gg\omega_f$, its magnitude approaches $K_D\omega_f$.
This bounded high-frequency gain is the classical motivation for the band-limited derivative feedback used in the main text.
With this parameter choice, the filtered controller satisfies $C_D(s)=G(s)$.
In the main-text model, the complete feedback response is $C(s)=G(s)$ without measurement filtering and $C(s)=G(s)H_{\rm meas}(s)=G_{\rm eff}(s)$ when it is included.

\section{Correlated detector noise}
\label{app:quantum_correlations}
\setcounter{equation}{0}

The uncorrelated-noise model above (imprecision $x_{\rm imp}$ and backaction force $F_{\rm ba}$ with vanishing cross-correlation) is sufficient to expose the basic quantum tradeoff, but it is not the most general quantum-limited detector.
In the linear-response description of continuous position measurement, the imprecision and backaction processes can be correlated, leading to an interference term in the closed-loop noise spectrum \cite{ClerkRMP2010}.
Such correlations arise, for example, in variational measurement strategies and more generally whenever the measured output quadrature is not orthogonal to the backaction-driving quadrature \cite{ClerkRMP2010,HabibiJOpt2016}.

\paragraph*{Noise spectra with correlations.}

We retain the same measurement record,
\begin{equation}
y(\omega)=x(\omega)+x_{\rm imp}(\omega),
\end{equation}
and define the (symmetrized) imprecision, backaction, and cross spectra $S_{xx}^{\rm imp}(\omega)$, $S_{FF}^{\rm ba}(\omega)$, and
\begin{equation}
S_{xF}(\omega)\equiv \frac12\left\langle x_{\rm imp}(\omega)\,F_{\rm ba}(-\omega)+F_{\rm ba}(-\omega)\,x_{\rm imp}(\omega)\right\rangle .
\end{equation}
For the correlated-detector model considered here, we impose the input-referred noise constraint
\begin{equation}
S_{xx}^{\rm imp}(\omega)\,S_{FF}^{\rm ba}(\omega)-|S_{xF}(\omega)|^2 \;\ge\; \frac{\hbar^2}{4\eta}.
\label{eq:app_corr_quantum_constraint}
\end{equation}
It is convenient to parameterize correlations by a dimensionless coefficient
\begin{equation}
\rho(\omega)\equiv \frac{S_{xF}(\omega)}{\sqrt{S_{xx}^{\rm imp}(\omega)\,S_{FF}^{\rm ba}(\omega)}},
\qquad |\rho(\omega)|\le 1,
\label{eq:rho_def}
\end{equation}
so that Eq.~\eqref{eq:app_corr_quantum_constraint} implies $|\rho(\omega)|^2 \le 1-\hbar^2/[4\eta S_{xx}^{\rm imp}(\omega)S_{FF}^{\rm ba}(\omega)]$.
For completeness, one may saturate Eq.~\eqref{eq:app_corr_quantum_constraint} near $\omega\simeq\omega_0$ to obtain a simple quantum-limited correlated-detector model.

\paragraph*{Closed-loop spectrum with an interference term.}

With the same band-limited derivative feedback filter $G_{\rm eff}(\omega)$ used in Sec.~\ref{sec:full_spectrum_bandwidth} (including the measurement-bandwidth filter if present), the closed-loop displacement spectrum generalizes to
\begin{align}
&S_{xx}(\omega)=|\chi_{\rm eff}(\omega)|^2\Big(
S_{FF}^{\rm th}(\omega)+S_{FF}^{\rm ba}(\omega)\nonumber\\&+|G_{\rm eff}(\omega)|^2 S_{xx}^{\rm imp}(\omega)
-2\,\mathrm{Re}\!\left[G_{\rm eff}(\omega)\,S_{xF}(\omega)\right]\Big),
\label{eq:Sxx_correlated}
\end{align}
where $\chi_{\rm eff}(\omega)=\chi_m(\omega)/[1+\chi_m(\omega)G_{\rm eff}(\omega)]$ as before.
The last term in Eq.~\eqref{eq:Sxx_correlated} has not been present before - depending on the sign and phase of $S_{xF}$ relative to $G_{\rm eff}$, correlations can \emph{either} reduce \emph{or} increase the net injected noise.
Consequently, correlated detectors can shift the optimal controller parameters and can partially mitigate the performance degradation at large derivative gain or cutoff identified in Fig.~\ref{fig:quantum_feedback_panels}.

\paragraph*{High-$Q$ implication.}

Under the near-resonant assumptions stated in Appendix~\ref{app:highQ_reduction}, Eq.~\eqref{eq:Sxx_correlated} adds a contribution \emph{linear} in $\gamma_{\rm fb}$ to the effective noise floor.
Accordingly, the approximate objective takes the generic form
\begin{equation}
n_{\rm eff}+\frac12 \approx \frac{A+B_n\gamma_{\rm fb}^2-C\gamma_{\rm fb}}
{\gamma_u+\alpha_d\gamma_{\rm fb}},
\label{eq:neff_highQ_correlated_form}
\end{equation}
where $A$ and $B_n$ reduce to Eq.~\eqref{eq:AB_phase_lag_defs} in the uncorrelated limit, and $C$ encodes the near-resonant correlation contribution through $\mathrm{Re}[G_{\rm eff}(\omega_0)S_{xF}(\omega_0)]$.

Equation~\eqref{eq:neff_highQ_correlated_form} makes explicit that correlations can shift the optimizer and lower the achievable minimum by introducing an interference term that competes with the quadratic reinjection term.

\section{High-$Q$ reduction}
\label{app:highQ_reduction}
\setcounter{equation}{0}

This appendix derives the approximate high-$Q$ expression used in Sec.~\ref{sec:phase_lag_bandwidth}. 
The purpose is to show how the force-noise terms and the feedback-induced damping enter the occupation estimate.

For this reduction, we assume that the feedback-broadened resonance remains narrow compared with $\omega_0$, that its frequency shift is small, and that it dominates both the displacement and momentum variance integrals.
A large intrinsic mechanical quality factor alone does not guarantee these conditions.
The effective susceptibility may therefore be approximated by a Lorentzian with total near-resonant amplitude-damping rate
\begin{equation}
\Gamma_{\rm eff} = \gamma_u + \alpha_d\gamma_{\rm fb},
\label{eq:app_Gamma_eff}
\end{equation}
where $\alpha_d$ is the phase-lag factor defined in Eq.~\eqref{eq:alpha_d}. 
The real part of the loop filter gives a small frequency shift in this narrowband estimate, while the imaginary part gives the viscous damping retained in Eq.~\eqref{eq:app_Gamma_eff}. 
The full-spectrum calculation in Sec.~\ref{sec:full_spectrum_bandwidth} retains the complete complex response.

For a two-sided symmetrized force spectrum that is approximately flat across the mechanical linewidth, the oscillator occupation satisfies
\begin{equation}
n_{\rm eff}+\frac12
\simeq
\frac{S_{FF}^{\rm tot}(\omega_0)}
{4m\hbar\omega_0\,\Gamma_{\rm eff}},
\label{eq:app_highQ_force_to_occupation}
\end{equation}
where $S_{FF}^{\rm tot}$ is the total force-noise spectrum acting on the oscillator near resonance. 
This convention is consistent with Eq.~\eqref{eq:thermal_force}: when feedback and backaction are absent, $S_{FF}^{\rm th}=4m\gamma_u\hbar\omega_0(n_{\rm th}+1/2)$ and $\Gamma_{\rm eff}=\gamma_u$, so Eq.~\eqref{eq:app_highQ_force_to_occupation} gives $n_{\rm eff}=n_{\rm th}$.

In the uncorrelated quantum-limited model, the near-resonant force noise is
\begin{equation}
S_{FF}^{\rm tot}(\omega_0)
=
S_{FF}^{\rm th}
+
S_{FF}^{\rm ba}
+
\left|G_{\rm eff}(\omega_0)\right|^2 S_{xx}^{\rm imp}.
\label{eq:app_total_force_noise_highQ}
\end{equation}
Using Eq.~\eqref{eq:alpha_n}, the imprecision-reinjection term is
\begin{equation}
\left|G_{\rm eff}(\omega_0)\right|^2 S_{xx}^{\rm imp}
=
\left(2m\gamma_{\rm fb}\omega_0\right)^2
\alpha_n S_{xx}^{\rm imp}.
\label{eq:app_imp_reinjection_highQ}
\end{equation}
Substituting Eqs.~\eqref{eq:thermal_force} and \eqref{eq:app_imp_reinjection_highQ} into Eq.~\eqref{eq:app_highQ_force_to_occupation} gives
\begin{equation}
n_{\rm eff}+\frac12
\approx
\frac{
\gamma_u\left(n_{\rm th}+\frac12\right)
+
\dfrac{S_{FF}^{\rm ba}}{4m\hbar\omega_0}
+
\dfrac{m\omega_0\alpha_n S_{xx}^{\rm imp}}{\hbar}\gamma_{\rm fb}^2
}
{\gamma_u+\alpha_d\gamma_{\rm fb}}.
\label{eq:app_highQ_neff_expanded}
\end{equation}
Thus,
\begin{equation}
n_{\rm eff}+\frac12
\approx
\frac{A+B_n\gamma_{\rm fb}^2}
{\gamma_u+\alpha_d\gamma_{\rm fb}},
\label{eq:app_highQ_neff_compact}
\end{equation}
with
\begin{equation}
A=
\gamma_u\left(n_{\rm th}+\frac12\right)
+\frac{S_{FF}^{\rm ba}}{4m\hbar\omega_0},
\quad
B_n=
\frac{m\omega_0\alpha_n S_{xx}^{\rm imp}}{\hbar}.
\label{eq:app_highQ_AB_defs}
\end{equation}
This is the high-$Q$ estimate used in Eq.~\eqref{eq:neff_highQ_phase_lag}; in the dimensionless map, the same expression is evaluated with $\alpha_d$ and $\alpha_n$ written in terms of $r_f$ and $r_{\rm meas}$.
It is intended as a compact design rule; the numerical full-spectrum calculation keeps the complete frequency dependence of $\chi_{\rm eff}(\omega)$ and $G_{\rm eff}(\omega)$.

\onecolumn
\section{Summary of notation}
\label{app:notation}
\setcounter{equation}{0}
\ifdefined\nolinenumbers\nolinenumbers\fi

\noindent\begin{minipage}{\textwidth}
\centering

\captionof{table}{Principal physical quantities, response functions and high-$Q$ coefficients. 
The bandwidth ratios and the factors $\alpha_d$ and $\alpha_n$ are dimensionless.
}
\label{tab:symbol_summary}

\renewcommand{\arraystretch}{1.1}

\begin{tabular}{@{}p{0.18\textwidth}
                  p{\dimexpr0.82\textwidth-2\tabcolsep\relax}@{}}
\toprule
Symbol & Definition \\
\midrule

\multicolumn{2}{@{}l@{}}{
\textit{Bath and occupation}
} \\[2pt]

$T$ &
Thermal-bath temperature. \\

$n_{\rm th}(T)$ &
Mean thermal occupation at temperature $T$, defined in Eq.~\eqref{eq:nth_def}. \\

$n_{\rm eff}$ &
Effective occupation defined from the mean mechanical energy in Eq.~\eqref{eq:neff_def}. \\

\midrule
\multicolumn{2}{@{}l@{}}{
\textit{Detector and noise}
} \\[2pt]

$\eta$ &
Detector efficiency, $0<\eta\le1$. \\

$S_{xx}^{\rm imp}$ &
Measurement-imprecision displacement-noise spectrum.$^{\dagger}$ \\

$S_{FF}^{\rm th}$ &
Thermal force-noise spectrum.$^{\dagger}$ \\

$S_{FF}^{\rm ba}$ &
Measurement-backaction force-noise spectrum.$^{\dagger}$ \\

$S_{xx}$ &
Closed-loop displacement spectrum.$^{\dagger}$ \\

\midrule
\multicolumn{2}{@{}l@{}}{
\textit{Feedback and bandwidth}
} \\[2pt]

$\gamma_{\rm fb}$ &
Feedback-gain parameter giving the near-resonant damping contribution $\alpha_d\gamma_{\rm fb}$. \\

$\omega_f$ &
Roll-off angular frequency of the derivative controller. \\

$\omega_{\rm meas}$ &
Angular-frequency bandwidth of the measurement response $H_{\rm meas}$. \\

$q_f$ &
Inverse controller-bandwidth ratio, $q_f=\omega_0/\omega_f$. \\

$q_{\rm meas}$ &
Inverse measurement-bandwidth ratio, $q_{\rm meas}=\omega_0/\omega_{\rm meas}$. \\

$r_f$ &
Normalized controller bandwidth, $r_f=\omega_f/\omega_0=1/q_f$. \\

$r_{\rm meas}$ &
Normalized measurement bandwidth, $r_{\rm meas}=\omega_{\rm meas}/\omega_0=1/q_{\rm meas}$. \\

$\alpha_d$ &
Near-resonant damping factor defined in Eq.~\eqref{eq:alpha_d}.
Positive values give damping-producing feedback. \\

$\alpha_n$ &
Near-resonant imprecision-noise reinjection factor defined in Eq.~\eqref{eq:alpha_n}. \\

\midrule
\multicolumn{2}{@{}l@{}}{
\textit{Response functions and high-$Q$ coefficients}
} \\[2pt]

$\chi_m$ &
Mechanical displacement susceptibility without feedback, defined in Eq.~\eqref{eq:classical_chi_m}. \\

$\chi_{\rm eff}$ &
Mechanical displacement susceptibility with feedback, including
the measurement response when present. \\

$G$ &
Band-limited derivative-controller response defined in
Eq.~\eqref{eq:band-limited_derivative}. \\

$H_{\rm meas}$ &
Measurement-bandwidth response defined in Eq.~\eqref{eq:Hmeas}. \\

$G_{\rm eff}$ &
Combined feedback response, $G_{\rm eff}=G H_{\rm meas}$. \\

$A$ &
Gain-independent thermal and backaction term in the high-$Q$ numerator, defined in Eq.~\eqref{eq:AB_phase_lag_defs}. \\

$B_n$ &
Coefficient of the imprecision-reinjection term $B_n\gamma_{\rm fb}^2$, defined in Eq.~\eqref{eq:AB_phase_lag_defs}. \\

\bottomrule
\end{tabular}

\par\smallskip
\begin{minipage}{\textwidth}
\footnotesize
$^{\dagger}$All listed noise spectra are two-sided and symmetrized, with angular frequency $\omega$ as the frequency variable.
\end{minipage}

\end{minipage}
\twocolumn

\newpage
\section{Full closed-loop stability}
\label{app:closed_loop_stability}
\setcounter{equation}{0}

The deterministic closed-loop poles follow from $1+\chi_m(s)G(s)H_{\rm meas}(s)=0$.
The feedback path is described by the effective response $G_{\rm eff}(s)=G(s)H_{\rm meas}(s)$, where $G(s)$ is the band-limited derivative controller and $H_{\rm meas}(s)$ describes the finite measurement bandwidth.
Substitution into the deterministic closed-loop equations gives the characteristic polynomial below.
We introduce the dimensionless variables
\begin{equation}
\lambda=\frac{s}{\omega_0},\qquad
\epsilon=\frac{\gamma_u}{\omega_0},\qquad
g=\frac{\gamma_{\rm fb}}{\omega_0},
\label{eq:app_RH_normalization}
\end{equation}
and
\begin{equation}
R=r_f+r_{\rm meas},\qquad P=r_f r_{\rm meas}.
\label{eq:app_RH_RP}
\end{equation}
The characteristic equation is $p(\lambda)=0$, where
\begin{equation}
\begin{split}
p(\lambda)={}&(\lambda^2+2\epsilon\lambda+1)
              (\lambda+r_f)(\lambda+r_{\rm meas})\\
             &+2gP\lambda .
\end{split}
\label{eq:app_RH_characteristic}
\end{equation}
Equivalently,
\begin{equation}
p(\lambda)=\lambda^4+a_1\lambda^3+a_2\lambda^2+a_3\lambda+a_4,
\label{eq:app_RH_quartic}
\end{equation}
with
\begin{equation}
\begin{aligned}
a_1&=R+2\epsilon,\\
a_2&=P+1+2\epsilon R,\\
a_3&=R+2P(\epsilon+g),\\
a_4&=P.
\end{aligned}
\label{eq:app_RH_coefficients}
\end{equation}
For $r_f,r_{\rm meas}>0$, $\epsilon>0$, and $g\ge0$, all coefficients are positive.
The remaining quartic Routh--Hurwitz conditions are
\begin{equation}
\begin{aligned}
\Delta_2&=a_1a_2-a_3>0,\\
\Delta_3&=a_3\Delta_2-a_1^2P>0.
\end{aligned}
\label{eq:app_RH_determinants}
\end{equation}
Since $a_3>0$ and $a_1^2P>0$, the condition $\Delta_3>0$ also implies $\Delta_2>0$.
The final inequality can be written as
\begin{equation}
a_3^2-a_1a_2a_3+a_1^2P<0,
\label{eq:app_RH_a3_quadratic}
\end{equation}
whose roots in $a_3$ are
\begin{equation}
a_{3,\pm}=\frac{a_1}{2}
\left(a_2\pm\sqrt{a_2^2-4P}\right).
\label{eq:app_RH_a3_limits}
\end{equation}
At $g=0$, the characteristic polynomial factorizes into the intrinsically damped oscillator and the two stable filter factors.
Thus $g=0$ lies strictly inside the stable interval when $\epsilon>0$.
Since $a_3$ increases linearly with $g$, the nonnegative-gain stability range is
\begin{equation}
0\le g<g_{\rm RH},
\label{eq:app_RH_stability_range}
\end{equation}
where
\begin{equation}
\begin{aligned}
g_{\rm RH}
={}&\frac{R+2\epsilon}{4P}
\left(a_2+\sqrt{a_2^2-4P}\right)\\
&-\frac{R}{2P}-\epsilon.
\end{aligned}
\label{eq:app_g_RH}
\end{equation}
Here $a_2=P+1+2\epsilon R$, so the limit depends only on $r_f$, $r_{\rm meas}$, and $\epsilon$.
Equality is excluded from asymptotic stability.
At the threshold, a conjugate pair lies on the imaginary axis, with normalized crossing frequency $\Omega_{\rm cross}=\sqrt{a_3/a_1}$ evaluated at $g=g_{\rm RH}$.

The condition $r_f r_{\rm meas}>1$ determines the sign of the feedback-damping contribution evaluated at the bare resonance, whereas Eq.~\eqref{eq:app_RH_stability_range} constrains all four closed-loop poles.
For positive intrinsic damping, sufficiently small gains can remain stable even when $r_f r_{\rm meas}\le1$.
Conversely, positive near-resonant feedback damping does not prevent instability at gains above $g_{\rm RH}$.
The numerical comparison in Sec.~\ref{sec:dimensionless_map} uses Eq.~\eqref{eq:app_g_RH} and independently checks the four roots at every evaluated high-$Q$ optimal gain.

\section{Full-spectrum moments and gain optimization}
\label{app:full_spectrum_moments}
\setcounter{equation}{0}

This evaluation uses the uncorrelated, frequency-independent noise spectra of the main text and finite positive filter bandwidths.
With $u=\omega/\omega_0$, define
\begin{equation}
\begin{aligned}
\sigma&=\frac{m\omega_0^2S_{xx}^{\rm imp}}{\hbar},
&J&=r_f^2+r_{\rm meas}^2,\\
q_{\rm th}&=4\epsilon\left(n_{\rm th}+\frac12\right),
&q_{\rm ba}&=\frac{1}{4\eta\sigma}.
\end{aligned}
\label{eq:full_moment_noise_parameters}
\end{equation}
Using $P$, $p(\lambda)$, and the quartic coefficients from Appendix~\ref{app:closed_loop_stability}, the normalized displacement spectrum $\mathcal S_X(u)=(m\omega_0^2/\hbar)S_{xx}(\omega_0u)$ is
\begin{equation}
\begin{aligned}
\mathcal S_X(u)={}&
\frac{(q_{\rm th}+q_{\rm ba})(u^4+Ju^2+P^2)}{|p(iu)|^2}\\
&+\frac{4\sigma g^2P^2u^2}{|p(iu)|^2}.
\end{aligned}
\label{eq:full_moment_spectrum}
\end{equation}
This expression is obtained by substituting $G(\omega)$ and $H_{\rm meas}(\omega)$ into Eq.~\eqref{eq:Sxx_with_bandwidth} and rewriting the result with the common denominator $|p(iu)|^2$.
The full frequency dependence is retained, without a near-resonant approximation.

For a strictly stable quartic, introduce
\begin{equation}
I_j=\frac{1}{\pi}\int_0^\infty
\frac{u^j}{|p(iu)|^2}\,du,\qquad j=0,2,4,6.
\label{eq:full_moment_integrals}
\end{equation}
In dimensionless time $\tau=\omega_0t$, let $z(\tau)$ be the output of the auxiliary system with transfer function $1/p(\lambda)$ driven by white noise of unit spectral intensity.
The state $(z,z',z'',z^{(3)})$ then has variances $(I_0,I_2,I_4,I_6)$, where primes denote derivatives with respect to $\tau$.
Its stationary covariance equations imply
\begin{equation}
\begin{aligned}
I_4&=a_2I_2-PI_0,& a_1I_4&=a_3I_2,\\
I_6&=a_2I_4-PI_2,& a_1I_6-a_3I_4&=\frac12.
\end{aligned}
\label{eq:full_moment_covariance_relations}
\end{equation}
Hence, with the Hurwitz determinants defined in Eq.~\eqref{eq:app_RH_determinants},
\begin{equation}
\begin{aligned}
I_0&=\frac{\Delta_2}{2P\Delta_3},
&I_2&=\frac{a_1}{2\Delta_3},\\
I_4&=\frac{a_3}{2\Delta_3},
&I_6&=\frac{a_2a_3-a_1P}{2\Delta_3}.
\end{aligned}
\label{eq:full_moment_closed_forms}
\end{equation}
Writing $q=q_{\rm th}+q_{\rm ba}$, the normalized physical moments are
\begin{equation}
\begin{aligned}
V_x&=\frac{m\omega_0\langle x^2\rangle}{\hbar}\\
   &=q(I_4+JI_2+P^2I_0)+4\sigma g^2P^2I_2,\\
V_p&=\frac{\langle p^2\rangle}{m\hbar\omega_0}\\
   &=q(I_6+JI_4+P^2I_2)+4\sigma g^2P^2I_4.
\end{aligned}
\label{eq:full_moment_physical_variances}
\end{equation}
The occupation is $n_{\rm eff}^{\rm full}=(V_x+V_p)/2-1/2$.
Thermal and backaction contributions can be evaluated separately by using $q_{\rm th}$ and $q_{\rm ba}$ in the force terms.

At fixed bandwidths and physical parameters, these expressions give
\begin{equation}
n_{\rm eff}^{\rm full}(g)+\frac12=
\frac{N(g)}{4\Delta_3(g)},
\label{eq:full_moment_rational_objective}
\end{equation}
where $N$ is cubic and $\Delta_3$ is quadratic in $g$.
At an interior stationary point, differentiation with respect to $g$ gives
\begin{equation}
N'(g)\Delta_3(g)-N(g)\Delta_3'(g)=0,
\label{eq:full_moment_stationary_condition}
\end{equation}
a polynomial equation of degree at most four.
For each evaluated bandwidth pair, real-root isolation with exact rational coefficients identifies every candidate in $0<g<g_{\rm RH}$.
We compare these candidates with $g=0$ and verify the divergent limit as $g\to g_{\rm RH}^{-}$ for the parameters used in the map.
The marginal endpoint is excluded.

These expressions evaluate the full-frequency moments of the constant-noise model without introducing a numerical frequency cutoff.
Independent calculations use the original four-state covariance equations and pole-resolved frequency quadrature on the samples stated in the main text.
For the normalized moments, we use a relative comparison tolerance of $10^{-5}$ and an absolute tolerance of $10^{-9}$.
For the occupation, the corresponding tolerances are $10^{-5}$ and $10^{-7}$.
These numerical tolerances are separate from the $10\%$ and $25\%$ approximation-error levels in Fig.~\ref{fig:dimensionless_map}\,(b).

\end{appendices}

\bibliography{sn-bibliography}

\end{document}